\documentclass[aip,sd, amsmath,amssymb, reprint]{revtex4-1}
\usepackage[british,UKenglish,USenglish,english,american]{babel}
\usepackage{graphicx}% Include figure files
\usepackage{dcolumn}% Align table columns on decimal point
\usepackage{bm}% bold math
\usepackage[pdftex]{color}
\usepackage{tikz}        % drawing toolbox
\usepackage{tikz}  
\usetikzlibrary{decorations.pathmorphing}
\usetikzlibrary{patterns}
\usepackage{xcolor} 
\usepackage{pgfplots}
\usetikzlibrary{shapes.geometric}
\usetikzlibrary{calc,3d}
\newcommand{\0} {\textbf{0}}

\renewcommand{\theequation}{\thesection.\arabic{equation}}

\newcommand {\Bc}  {\mathcal{B}}

\newcommand {\Vc}  {\mathcal{V}}

\newcommand {\eb} {\mathbf{e}}

\newcommand {\mb} {\mathbf{m}}
\newcommand {\nb} {\mathbf{n}}

\newcommand {\tb} {\mathbf{t}}

\newcommand {\Bb} {\mathbf{B}}
\newcommand {\Cb} {\mathbf{C}}

\newcommand {\Eb} {\mathbf{E}}
\newcommand {\Fb} {\mathbf{F}}

\newcommand {\Ib} {\mathbf{I}}

\newcommand {\Sb} {\mathbf{S}}
\newcommand {\Tb} {\mathbf{T}}

\renewcommand{\r} {\textcolor{black}}
\renewcommand{\b} {\textcolor{black}}
\xdefinecolor{myblue}{rgb}{.20,.20,.70}
\xdefinecolor{mygreen}{RGB}{0, 105, 0}
\xdefinecolor{mygreen}{RGB}{0, 155, 0}
\xdefinecolor{greenfibers}{RGB}{0, 65, 0}
\begin{document}

\preprint{AIP/123-QED}

\title[Anisotropic stroke--curves]{Actuation performances of anisotropic gels}% Force line breaks with \\
%\thanks{Footnote to title of article.}

\author{P. Nardinocchi}
\email{paola.nardinocchi@uniroma1.it}
% \affiliation{Dipartimento di Ingegneria Strutturale e Geotecnica, Sapienza Universit\`a di Roma.}%Lines break automatically or can be forced with \\
\affiliation{Dipartimento di Ingegneria Strutturale e Geotecnica, Sapienza Universit\`a di Roma, via Eudossiana 18, 00184 ROMA}%
\author{L. Teresi}
\email{teresi@uniroma3.it}
 %\homepage{http://www.Second.institution.edu/~Charlie.Author.}
\affiliation{Dipartimento di Matematica e Fisica, Universit\`a Roma Tre, via della Vasca Navale 84, 00146, ROMA%
%\\This line break forced% with \\
}%

\date{\today}% It is always \today, today,

\begin{abstract}
We investigated the actuation performances of anisotropic gels driven by mechanical and chemical stimuli, in terms of both deformation processes and stroke--curves, and distinguished between the fast response of gels before diffusion starts and the asymptotic response attained at the steady state. We also showed as the range of forces that an anisotropic hydrogel can exert when constrained is especially wide;indeed, changing fiber orientation allows to induce shear as well as transversely isotropic extensions.
\end{abstract}

%Use showkeys class option if keyword
\maketitle

%\begin{quotation}
%The ``lead paragraph'' is encapsulated with the \LaTeX\ 
%\verb+quotation+ environment and is formatted as a single paragraph before the first section heading. 
%(The \verb+quotation+ environment reverts to its usual meaning after the first sectioning command.) 
%Note that numbered references are allowed in the lead paragraph.
%%
%The lead paragraph will only be found in an article being prepared for the journal \textit{Chaos}.
%\end{quotation}

%
\section{Introduction}
Soft active materials admit deformations and displacements that can be triggered through a wide range of external stimuli such as electric field, pH, temperature, solvent absorption.\cite{Dai2009,Kim2012SM,JAP2013,Pezzulla:2015} The effectiveness of these systems may critically depend on the capability of achieving both prescribed changes in shape and size, and on the range of performances in actuator applications, which can 
involve isotropic or fibrous gels. We recently presented an investigation about 
\b{fiber reinforced gels, a soft composite material, whose shape changes can be programmed by adjusting both fibers orientation and stiffness, and about the possible shape changes that they realize in free--swelling conditions;
we also discussed the role of fibres's orientation  in determining  these shape changes.\cite{JAP2015}  
Moreover, we showed as, by considering geometric composites made of homogeneous layers of fibrous gels, an even larger variety of shapes can be generated starting from a flat strip.\cite{SM2015}}

Fibrous or anisotropic gels  are at the center of many recent realisation dealing with fibrous soft material inspired by plant world,\cite{Osorio2012} natural filtration systems,\cite{Liu2015_SM} biomedical materials for cardiovascular medicine,\cite{Millon2006}  polymer hydrogels with anisotropies in structure and optical properties, \cite{Miyamoto2013} as well as theoretical investigations. These different applications exploit the ability of such gels to undergo anisotropic swelling and to show an anisotropic mechanical response; both of these elements characterise the mechanics of fibrous hydrogels. 

To the best of our knowledge, the performances of fibrous hydrogels in actuator applications  have not been extensively studied.\cite{Liu2015} As it is well known, actuators are usually characterized by their force--stroke curves, which deliver critical information when designing an actuator.\cite{Cai2010,Illeperuma2013} In particular, the range of forces that an anisotropic hydrogel can exert when constrained is especially wide. Indeed, changing fiber orientation in gel cubic elements allows to induce shear as well as transversely isotropic extensions, under free--swelling conditions.\cite{JAP2015} Correspondingly, under appropriate imposed deformations, \textit{i.e.} boundary constraints, anisotropic gels can exert  both  tangential and normal forces, depending on fiber orientation. Moreover, due to swelling, such forces decrease from the instantaneous values attained before diffusion starts to the lower values attained at the steady state.\cite{Urayama:2012} 

This paper aims to investigate performances of anisotropic gels driven by mechanical and chemical stimuli, in terms of both deformation processes and stroke--curves, and distinguish between the fast response of gels before diffusion starts and the asymptotic response attained at the steady state. We start from a Background section devoted to revisit a few results related to isotropic gel actuators. Then, with reference to a thermodynamic model which can be viewed as the extension of the well--known Flory--Rehner model, a few prototypical anisotropic actuators are investigated, and the corresponding deformation processes and stroke--curves are discussed.  
%
%%%%%%%%%%%%%%%%%%%%%%%%%%%%%%%%%%%%%%%%%%%%%%%%%%%%%%%%%%
%
\section{Background}
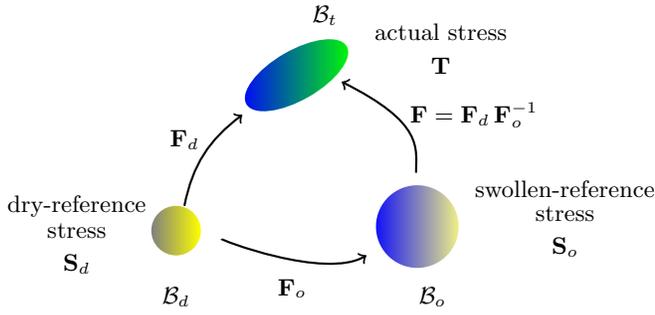
\begin{figure}\centering
\begin{tikzpicture}[scale=1.1]
   % Grad p
   \draw[->][thick] (0.1,0.3) .. controls (0.2,0.8) and (0.4,1.1) .. (0.8,1.4);
   \shade[left color=gray,right color=yellow] (0,0) ellipse (0.3 and 0.3);
   % G
   \draw[->][thick] (0.55,-0.1) .. controls (0.5,-0.1) and (1.8,-0.6) .. (2.3,-0.3);
   \shade[left color=blue,right color=yellow!50, rotate=-30] 
   (2.5,1.5) ellipse (0.5 and 0.5);
   % F=Grad p G^-1
   \draw[->][thick] (2.9,0.7) .. controls (2.9,1) and (3.,1.3) .. (2.,1.8);
   \shade[left color=blue,right color=green, rotate=30] 
   (2.2,0.9) ellipse (0.7 and 0.3);
   % Text
   \draw (3.1,-0.80) node {$\Bc_o$} (4.7,0.5) node {swollen-reference}
                                  (4.7,0.2) node {stress}
                                  (4.7,-0.2) node {$\Sb_o$};
   \draw (1.4,-0.70) node {$\Fb_o$};
   \draw (0,-0.80) node {$\Bc_d$} (-1.2,0.3) node {dry-reference}
                                  (-1.2,0.) node {stress}
                                  (-1.2,-0.4) node {$\Sb_d$};
   \draw (0.1,1.1) node {$\Fb_d$};
   \draw (1.8,2.6) node {$\Bc_t$} (3.2,2.4) node {actual stress}
                                  (3.2,2.) node {$\Tb$};
   \draw (3.6,1.4) node {$\Fb=\Fb_d\,\Fb_o^{-1}$};
\end{tikzpicture}
\caption{\label{fig:1} Given a dry ball $\Bc_d$, the swollen and stress--free  state $\Bc_o$ only differs for a change in size, whereas the actual state $\Bc_t$ for a change in size and possibly in shape.}
\end{figure}

Our starting point is the multiphysics model presented and discussed in Ref. \citenum{JMPS2013}; therein, three different states of a gel body were introduced: a dry state $\Bc_d$, a swollen and stress--free state $\Bc_o$,  and an actual state $\Bc_t$ (see cartoon in figure \ref{fig:1}). 
Then, the constitutive equation for the stress $\Sb_d$ ($[\Sb_d]$=Pa = J/m$^3$) at the dry configuration $\Bc_d$, from now on denoted as dry--reference stress,  
and for the chemical potential $\mu$ ($[\mu]$=J/mol) were derived from the classical Flory--Rehner thermodynamic context. The Flory--Rehner model\cite{FR1,FR2} for stress diffusion in gels is based on a free energy $\psi$  per unit dry  volume which depends on the deformation gradient $\Fb_d$ from the initial dry configuration of the polymer gel through an elastic component $\psi_e$, and on the molar solvent concentration $c_d$ per unit dry volume ($[c_d]=$mol/m$^3$) through a polymer--solvent mixing energy $\psi_m$: $\psi=\psi_e+\psi_m$. We include in the definition of the free--energy a volumetric constraint prescribing that changes in volume are only due to solvent absorption or release. In order to account for such a constraint, we relax the Flory--Rehner free energy by adding a term which enforces that constraint and write: 
\begin{equation}\label{isoFR}
\psi_r(\Fb_d,c_d,p) = \psi_e(\Fb_d) + \psi_m(c_d) -p(J_d-\hat J(c_d))\,.
\end{equation}
The pressure $p$ represents the reaction to the volumetric constraint, 
\b{which maintains the volume change due to the displacement equal to the one due to solvent absorption or release:} 
\begin{equation}\label{vol_const}
\b{J_d = \textrm{det}\,\Fb_d = \hat J(c_d) = 1 +\Omega\,c_d\,,}
\end{equation}
being $\Omega$ ($[\Omega]=$m$^3$/mol) the solvent molar volume. The function $\psi_r$ is called the Lagrangian function associated to the energy $\psi$, while $p$ is the Lagrangian multiplier which measures the sensitivity of the minimum energy to a change in the constraint. 
Key features of $\psi$ (or $\psi_r$) are the followings: (\textbf{i}) $\psi$ is a density per unit volume of the dry polymer; (\textbf{ii}) the elastic contribution $\psi_e$ hampers swelling; (\textbf{iii}) the mixing contribution $\psi_m$ favors swelling.   

The constitutive equation for the stress $\Sb_d$   
and the chemical potential $\mu$ ($[\mu]$=J/mol) come from thermodynamic issues and prescribe that 
\begin{equation}\label{csteqn}
\b{\Sb_d=\Sb_d(\Fb_d)-p\,\Fb_d^\star\,}
\quad\textrm{and}\,\quad
\mu=\mu(c_d)+p\,\Omega\,,
\end{equation}
with
\begin{equation}\label{mu_d}
\Sb_d(\Fb_d)=\frac{\partial\psi_e}{\partial\Fb_d}\quad\textrm{and}\quad\mu(c_d) = \frac{\partial\psi_m}{\partial c_d}\,.
\end{equation}
\b{The Flory--Rehner thermodynamic model prescribes a neo-Hookean  elastic energy $\psi_e$:}
\begin{equation}\label{nH}
\psi_e(\Fb_d) = \frac{G}{2}(\Fb_d\cdot\Fb_d-3)\,,
\end{equation}
\b{being $G$ the shear modulus of the dry polymer;
moreover, it prescribes the following polymer--solvent mixing energy:} 
\begin{equation}\label{FRm1}
\psi_m(c_d)= \frac{RT}{\Omega}\,h(c_d)\,,
\end{equation}
with
\begin{equation}\label{FRm2}
\b{h(c_d)=\Omega\,c_d\,\textrm{log}\frac{\Omega\,c_d}{1+\Omega\,c_d} + \chi\,\frac{\Omega\,c_d}{1+\Omega\,c_d}\,,
\quad [h]=1\,,}
\end{equation}
being $R$ ($[R]= $J/(K\,mol), $T$ ($[T]=$ K), and $\chi$ the universal gas constant, the temperature, and the Flory parameter, respectively. From \eqref{mu_d}$_1$ and \eqref{nH},  we derive the  constitutive equation $\Sb_d(\Fb_d)$ for the dry--reference stress; from \eqref{mu_d}$_2$, \eqref{FRm1}, and \eqref{FRm2} we derive the  constitutive equation $\mu(c_d)$ for the chemical potential. %
This latter can also be rewritten as function of $J_d$ \b{by exploiting the the volumetric constraint \eqref{vol_const}; with a slight abuse of notation, we write $\mu(c_d)=\mu(J_d)$:}
\begin{equation}\label{mud}
\mu(J_d)=RT\Bigl(\textrm{log}\frac{J_d-1}{J_d} + \frac{1}{J_d} + \frac{\chi}{J_d^2}\Bigr)\,.
\end{equation}
Performances of gels in terms of both deformation processes and stroke--curves driven by mechanical and chemical stimuli can be studied solving a time--dependent stress--diffusion problem based on appropriate balance equations and constitutive prescriptions for $\Sb_d$, $\mu$, and the solvent flux.\cite{JMPS2013,JAP2015} However, sometimes  homogeneous solutions are of interest, corresponding to steady states and/or, on the opposite side, to before--diffusion--starts states. A typical example deals with a gel body embedded into a solvent bath of assigned chemical potential $\mu_e$. In this case, the homogeneous solutions of the problem can be completely determined by data prescribed on the boundary in terms of boundary loads and/or constraints, and $\mu_e$. 

The simplest \b{example of such problems is the one with zero boundary loads}. In this case, mechanical and chemical balance laws prescribe $\Sb_d=\0$ and $\mu_e=\mu_o$, that is,  the swollen and stress--free state $\Bc_o$,
attained from $\Bc_d$ with $\Fb_d=\Fb_o=\lambda_o\Ib$, 
is completely defined by the value $\mu_o$ of the bath's chemical potential. The condition of zero stress yields the pressure $p$
\begin{equation}\label{stress0}
\b{G\Fb_o-p\,\Fb_o^\star=\0\,\,}
\quad\Rightarrow\quad
p=\frac{G}{\lambda_o}\,.
\end{equation}
By substituting $p$ in the constitutive relation for the chemical potential
yields a non linear equation relating $\mu_o$ and $\lambda_o$
\footnote
{We note that by considering the atmospheric pressure $p_a$ acting on $\Bc_o$, 
a correction factor must be added to the potential $\mu_o$:
equation \eqref{eqn:0} rewrites as
\begin{equation}\label{eqn:4}
 \mu(J_o)+\frac{G}{\lambda_o}\,\Omega = \mu_o-p_a\Omega\,;
\end{equation}
being $\Omega\simeq 10^{-5}$ m$^3$/mol and  $p_a\simeq 10^5$ Pa, 
the extra term to be added to $\mu_o$ is $p_a\,\Omega\simeq 1$ J/mol. 
}
\begin{equation}\label{eqn:0}
\mu(J_o)+\frac{G}{\lambda_o}\,\Omega = \mu_o\,,
\quad\textrm{with}\quad J_o=\lambda_o^3.
\end{equation}
For large deformation ($1/J_o\to 0$), equation \eqref{eqn:0} can be approximated, by estimating the leading order term in the Maclaurin asymptotic expansion in $1/J_o$, as\footnote{Known both $G$ and $\Omega$, an experimental setting with $\mu_o$ as control parameter allows to measure $\lambda_o$ and, from \eqref{eqcap}, the Flory parameter $\chi$.\citep{Urayama:2012}}
\begin{equation}\label{eqcap}
\frac{RT}{\Omega}(\chi-1/2)=\frac{\mu_o}{\Omega}J_o^2 -\frac{G}{\lambda_o}\,J_o^2\,.
\end{equation}
\b{In some cases, it may be convenient to use the free-swollen state $\Bc_o$  as reference configuration;
the deformation from $\Bc_o$ to the actual state $\Bc_t$ is then described by the deformation gradient 
$\Fb=\Fb_d\,\Fb_o^{-1}$.
The actual (Cauchy) stress $\Tb$ can then be represented in terms of the dry--reference stress $\Sb_d$, or
of the swollen--reference stress $\Sb$, defined as the 
\textit{push--forward} of  $\Sb_d$ and/or the \textit{pull--back} of $\Tb$:}
\begin{equation}\label{pufb}
\Sb=\underbrace{\frac{1}{J_o}\Sb_d\Fb_o^T }_{\textit{push--forward}}= \underbrace{\Tb\Fb^\star}_{\textit{pull--back}}\,,
\quad
\Tb=\frac{1}{J_d}\Sb_d\Fb_d^T\,.
\end{equation}
Using equations \eqref{csteqn}$_1$, \eqref{mu_d}$_1$, \eqref{nH}, and defined $\Bb_d=\Fb_d\,\Fb_d^T$, we have:
\begin{equation}\label{eqn:2}
\Sb=\frac{G}{J_o}\Fb\Fb_o\Fb_o^T -p\Fb^\star
\quad\textrm{and}\quad
\Tb=\frac{1}{J_d}\,G\,\Bb_d -p\Ib\,.
\end{equation}
\b{The deformation $\Fb$ can have different characteristics; in the following, we quickly revise two typical problems partially studied in Literature for isotropic gels.}
\subsection{Isotropic gels under step  pressure and dilation}
\b{Given a free-swollen state $\Bc_o$, we consider the steady--state $\Bc_t$ determined by 
the bath's chemical potential $\mu_{e}$, and by the external pressure $p_e$.
This state is described by the deformation field  $\Fb=\lambda\Ib$, and
we look for homogeneous solutions of the stress--diffusion problem, with boundary conditions}
\begin{equation}\label{eqn:1}
\Tb\,\nb=-p_e\,\nb\,\quad\textrm{and}\quad
\mu=\mu_e\,
\quad
\textrm{on}\,\partial\Bc_t\,,
\end{equation}
with $\nb$ the unit normal to 
$\partial\Bc_t$. Mechanical and chemical balances prescribe the spherical component $\sigma$ of the stress $\Tb=\sigma\Ib$ and the chemical potential within the gel:
\begin{equation}\label{eqnBAL}
\sigma=-p_e\quad\textrm{and}\quad \mu=\mu_e\quad\quad
\textrm{in}\,\Bc_t\,.
\end{equation}
Being $\Fb_o=\lambda_o\Ib$ and $\Fb=\lambda\Ib$, from equations \eqref{csteqn}$_1$, \eqref{mu_d}$_1$, \eqref{nH}, and \eqref{pufb}$_2$, \b{we obtain $\sigma$:}
\begin{equation}\label{eqn:1bis}
\sigma = \frac{G}{\lambda_o\,\lambda} -p \,.
\end{equation}
With this, equation (\ref{eqnBAL})$_1$ relates the (osmotic) 
pressure $p$ to the external pressure $p_e$ \b{and to the additional} deformation
$\lambda$ as:
\begin{equation}\label{pressioni}
p=\frac{G}{\lambda_o\;\lambda} + p_e\,.
\end{equation}
We focus on the slow response of the gel and assume that solvent migration  has reached a steady state. 
%
%Following Ref. \citenum{Urayama:2012}, just before diffusion starts the chemical balance \eqref{eqnBAL}$_2$ is still satisfied with $J=1$. Then, under the hypothesis $1/J_o\to 0$ and for $\mu_e=\mu_o=0$, equations \eqref{csteqn} and \eqref{pressioni} deliver 
%%
%\begin{equation}\label{creep2}
%\frac{G}{\lambda_o}\Bigl(1 - \frac{1}{\lambda}\Bigr) = p_e\,,
%\end{equation}
%%
%with $p_e$, measuring both the isotropic before--diffusion--start stress $\sigma_{bd}$ within the body and the intensity of the normal boundary traction (see equations \eqref{eqn:1}$_1$ and \eqref{eqnBAL}$_1$). From  equation \eqref{creep2}, $\lambda<1$ results in  $\sigma>0$, that is, solvent is expelled; on the contrary, $\lambda>1$ results in  $\sigma<0$, that is, solvent is absorbed.
% 
The characteristics of this response are determined by the equation (\ref{eqnBAL})$_2$ which, together with equations  \eqref{csteqn}$_2$ and \eqref{mud}, yields an implicit relation between the triplet $(p_e,\lambda,\mu_e)$: 
\begin{equation}\label{eqn:3}
\mu(J\,J_o)+\frac{G\,\Omega}{\lambda_o\,\lambda} +p_e\,\Omega=\mu_e\,,
\quad
 J\,J_o = (\lambda\,\lambda_o)^3\,.
 \end{equation}
Fixed the pair $(p_e,\mu_e)$, the stretch $\lambda$  determines the size of $\Bc_t$; alternatively, fixed the pair $(\lambda,\mu_e)$ with $\lambda$ an imposed dilation, $p_e$ determines both the isotropic  stress  within the body and the intensity of the normal boundary traction (see equations \eqref{eqn:1}$_1$ and \eqref{eqnBAL}$_1$).
\begin{figure}[h]
\centering
\includegraphics[scale=.6]{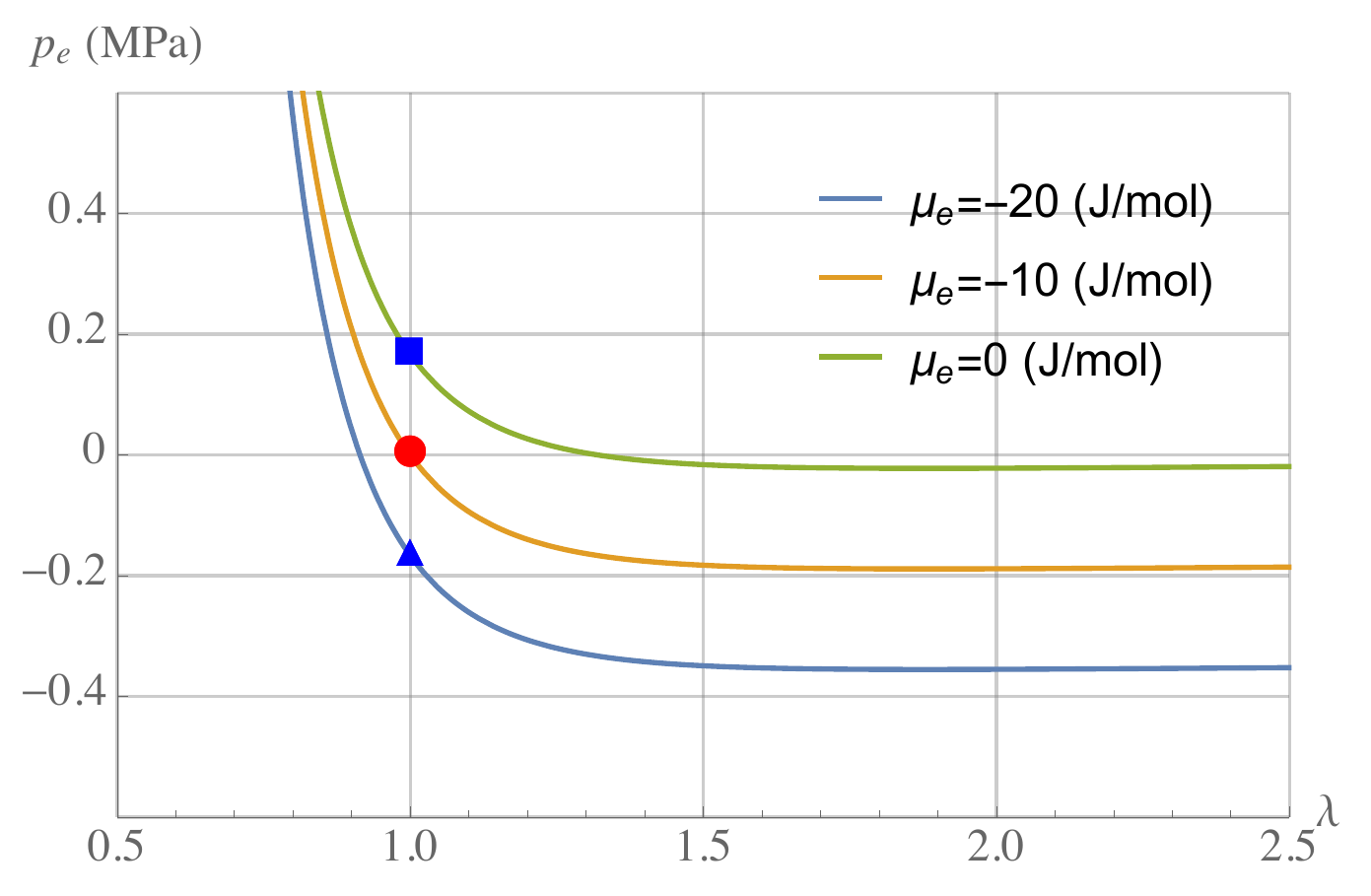}
\caption{Pressure--stroke curves $p_e(\lambda,\mu_e,\lambda_o)$ for three  
values of $\mu_e$ with $\lambda_o=2$, corresponding to $\mu_o=-10$ J/mol. When $\mu_e<\mu_o$ ($\mu_e>\mu_o$) the corresponding blocking pressure is determined on the vertical axis $\lambda=1$ in correspondence of the blue triangle (square) on the blue (green) curve.}
\label{figF:2}
\end{figure}
%
%Equations \eqref{creep2} and \eqref{creep1} have already been described in Ref. \citenum{Urayama:2012}, in the case of gels under uniaxial traction or extension. The decay in stress from $\sigma_{bd}$ to $\sigma_\infty$ when a gel into a solvent bath is subjected to a step dilation $\lambda$ is shown in figure  \ref{fig:2v5} in terms of degree of total reduction $\Delta\sigma=(\sigma_{bd}-\sigma_\infty)/\sigma_{bd}$.
%%
%\begin{figure}[h]
%\centering
%\includegraphics[scale=.5]{figure_2v5}
%\caption{Stress relaxation: difference between red and green solid lines measures the difference at different values $\lambda=\lambda_c$ between a fast response (value attained by the stress before diffusion starts) and a slow response (value attained by the stress at the steady state).}
%\label{fig:2v5}
%\end{figure}
%%

We may look for the pressure $p_e$  needed to keep a fixed dilation $\lambda$, under different $\mu_e$; in this case, equation \eqref{eqn:3}$_1$ can be recast as a function delivering $p_e$ in terms of $\lambda$ and $\mu_e$, with $\lambda_o$ (or, equivalently, $
\mu_o$) as a parameter:
\begin{equation}\label{eqn:3bis}
p_e = p_e(\lambda, \mu_e,\lambda_o)
    = \frac{1}{\Omega}\,(\mu_e - \mu(\lambda^3\,\lambda_o^3))
    - \frac{G}{\lambda_o\,\lambda}\,.
 \end{equation}
\b{Fixed the initial free-swollen conditions determined by $\mu_o$ (or, through (\ref{eqn:0}), by $\lambda_o$), 
we can have different stroke--curves $p_e$ versus $\lambda$, 
which depend on the new value of $\mu_e$;
figure \ref{figF:2} shows some of these curve for} $\mu_o=-10$ J/mol (and correspondingly, $\lambda_o=2$), and $\mu_e=\mu_o\pm 10$ J/mol, once fixed $\Omega=6\cdot 10^{-5}$m$^3$/mol, $\chi=0.2$, and $G=0.1$MPa. 
At different values of $\mu_e=-20, -10, 0$ J/mol, pressure--stroke curves intercept the axis $p_e=0$ at different values of $\lambda$ which correspond to free--swelling stretches.  
\b{For $\mu_e=\mu_o=-10$ J/mol, we recover the free-swollen reference state, that is $\lambda=1$, and $p_e=0$.}

In particular, equation \eqref{eqn:3bis} also allows to discuss the existence of a blocking pressure, that is, a pressure $p_e^\star$ which, \b{depending on the value of $\mu_e$}, 
\b{maintains $\lambda=1$, that is $\Bc_t=\Bc_o$.} 
\b{Equations  (\ref{pressioni})--(\ref{eqn:3})  with $\lambda=1$}
\begin{equation}\label{blockp}
\frac{G}{\lambda_o} - p=-p_e^\star\quad\textrm{and}\quad
\mu(J_o)+\frac{G\Omega}{\lambda_o}+p_e^\star\Omega=\mu_e\,,
\end{equation}
\b{characterise the blocking pressure $p_e^*$:} 
\begin{equation}\label{bp0}
p_e^\star = \frac{1}{\Omega}\;(\mu_e-\mu_o)\,.
\end{equation}
As expected, equation \eqref{bp0} \b{gives a null blocking pressure for $\mu_e=\mu_o=-10$ J/mol;
increasing (decreasing) $\mu_e$ requires a positive (negative) pressure $p_e^\star$ to maintain $\lambda=1$.}

It is worth noting that for large swelling--induced deformations, \textit{i.e.} $1/J_o\to 0$ and  $1/J\to 0$, equation \eqref{eqn:3} can be approximated as
\begin{equation}\label{creep1}
\frac{G}{\lambda_o}\Bigl(\frac{1}{\lambda^6} - \frac{1}{\lambda}\Bigr) = p_e\,,
\end{equation} 
where we set $\mu_o=\mu_e=0$.
\subsection{Isotropic gels under step traction and extension}
% 
%
% =======================================================
% FIGURE 
% =======================================================
\begin{figure*}[t]
\centering
\begin{tikzpicture}[x  = {(-0.5cm,-0.5cm)},
                    y  = {(0.9659cm,-0.25882cm)},
                    z  = {(0cm,1cm)},
                    scale = 0.4,
                    color = {mygreen}]
% style of faces
\tikzset{facestyle/.style={fill=mygreen,draw=greenfibers,opacity=.8,very thin,line join=round}}
% face "back" 
\begin{scope}[canvas is zy plane at x=0]
  \path[facestyle] (0,0) rectangle (1,1);
\end{scope}
% face  "left"
\begin{scope}[canvas is zx plane at y=0]
  \path[facestyle] (0,0) rectangle (1,1);
\end{scope}
% face "front"
\begin{scope}[canvas is zy plane at x=1]
  \path[facestyle] (0,0) rectangle (1,1);
\end{scope}
% face  "right"
\begin{scope}[canvas is zx plane at y=1]
  \path[facestyle] (0,0) rectangle (1,1);
  \shade[right color=green, left color=mygreen,opacity=.8] (0,0) rectangle (1,1);
\end{scope}
%face top
\begin{scope}[canvas is yx plane at z=1]
  \shade[left color=mygreen,right color=green,opacity=.8] (0,0) rectangle (1,1);
\end{scope}
%   
% REFERENCE FRAME
\begin{scope}[canvas is yx plane at z=0]
\draw[black,->, very thick] (0,1) -- (1,1) node [below] {\large{$2$}};
%\draw[black,->, very thick] (0,0) --(1,0); 
\end{scope}
\begin{scope}[canvas is zx plane at y=0]
\draw[black,->, very thick] (0,1) -- (1,1) node[left] {\large{$3$}};
%\draw[black,->, very thick] (0,0) -- (1,0);
\end{scope}
\begin{scope}[canvas is zx plane at y=0]
\draw[black,->, very thick] (0,1) -- (0,2) node[left] {\large{1}};
\end{scope}
\draw[black] (0,0,-3.) node[below]{a) Dry reference};
\draw[black] (0,0,4.) node[below]{$J_d=1$};
\end{tikzpicture}
%
%=============================
\hskip0.25cm %swollen
%=============================
%
\begin{tikzpicture}[x  = {(-0.5cm,-0.5cm)},
                    y  = {(0.9659cm,-0.25882cm)},
                    z  = {(0cm,1cm)},
                    scale = 0.4,
                    color = {mygreen}]
% style of faces
\tikzset{facestyle/.style={fill=mygreen,draw=greenfibers,opacity=.6,very thin,line join=round}}
% face "back" 
\begin{scope}[canvas is zy plane at x=0] \path[facestyle] (0,0) rectangle (2.64,2.64); \end{scope}
% face  "left"
\begin{scope}[canvas is zx plane at y=0] \path[facestyle] (0,0) rectangle (2.64,2.64); \end{scope}
% face "front"
\begin{scope}[canvas is zy plane at x=2.64] \path[facestyle] (0,0) rectangle (2.64,2.64); \end{scope}
% face  "right"
\begin{scope}[canvas is zx plane at y=2.64]
  \path[facestyle] (0,0) rectangle (2.64,2.64);
  \shade[right color=green, left color=mygreen,opacity=.8] (0,0) rectangle (2.64,2.64);
\end{scope}
%face top
\begin{scope}[canvas is yx plane at z=2.64]
  \shade[left color=mygreen,right color=green,opacity=.8] (0,0) rectangle (2.64,2.64);
\end{scope}
\draw[black] (0,0,-2.9) node[below]{b) Swollen \& unloaded};
\draw[black] (0,-1,4.4) node[below]{$J_d=J_o=\lambda_o^3$};
\end{tikzpicture}
%=============================
\hskip0.25cm  %traction istantaneous
%=============================
%
\begin{tikzpicture}[x  = {(-0.5cm,-0.5cm)},
                    y  = {(0.9659cm,-0.25882cm)},
                    z  = {(0cm,1cm)},
                    scale = 0.4,
                    color = {mygreen}]
% style of faces
\tikzset{facestyle/.style={fill=mygreen,draw=greenfibers,opacity=.6,very thin,line join=round}}
% face "back" 
\begin{scope}[canvas is zy plane at x=0] \path[facestyle] (0,0) rectangle (2.21,2.21); \end{scope}
% arrow back
\begin{scope}[canvas is xz plane at y=2.21/2]
\draw[black,->,ultra thick] (0,2.21/2) -- (-4,2.21/2) node[right] {\Large $\sigma_e$};
\end{scope}
% face  "left"
\begin{scope}[canvas is zx plane at y=0] \path[facestyle] (0,0) rectangle (2.21,3.76); \end{scope}
% face "front"
\begin{scope}[canvas is zy plane at x=3.76] \path[facestyle] (0,0) rectangle (2.21,2.21); \end{scope}
% arrow front 
\begin{scope}[canvas is xz plane at y=2.21/2]
\draw[black,->,ultra thick] (3.76,2.21/2) -- (7,2.21/2) node[left] {\Large $\sigma_e$};
\end{scope}
% face  "right"
\begin{scope}[canvas is zx plane at y=2.23]
  \path[facestyle] (0,0) rectangle (2.21,3.76);
  \shade[right color=green, left color=mygreen,opacity=.8] (0,0) rectangle (2.21,3.76);
\end{scope}
%face top
\begin{scope}[canvas is yx plane at z=2.21]
  \shade[left color=mygreen,right color=green,opacity=.8] (0,0) rectangle (2.21,3.76);
\end{scope}
\draw[black] (0,0,-5.)  node {c) Swollen \& loaded};
\draw[black] (0,0,-6.5) node {(fast response)};
\draw[black] (0,-2,3.4)  node{$J_d=J_o$};
\end{tikzpicture}
%
%=============================
\hskip0.25cm %traction steady response 
%=============================
%
%
\begin{tikzpicture}[x  = {(-0.5cm,-0.5cm)},
                    y  = {(0.9659cm,-0.25882cm)},
                    z  = {(0cm,1cm)},
                    scale = 0.4,
                    color = {mygreen}]
% style of faces
\tikzset{facestyle/.style={fill=mygreen,opacity=.6,very thin,line join=round}}
% face "back" 
\begin{scope}[canvas is zy plane at x=0] \path[facestyle] (0,0) rectangle (2.35,2.35); \end{scope}
% arrow back
\begin{scope}[canvas is xz plane at y=2.35/2]
\draw[black,->,ultra thick] (0,2.35/2) -- (-4,2.35/2) node[right] {\Large $\sigma_e$};
\end{scope}
% face  "left"
\begin{scope}[canvas is zx plane at y=0] \path[facestyle] (0,0) rectangle (2.35,4.13); \end{scope}
% face "front"
\begin{scope}[canvas is zy plane at x=4.13] \path[facestyle] (0,0) rectangle (2.35,2.35); \end{scope}
% arrow front 
\begin{scope}[canvas is xz plane at y=2.35/2]
\draw[black,->,ultra thick] (3,2.35/2) -- (7,2.35/2) node[left] {\Large $\sigma_e$};
\end{scope}
% face  "right"
\begin{scope}[canvas is zx plane at y=2.35]
  \path[facestyle] (0,0) rectangle (2.35,4.13);
  \shade[right color=green, left color=mygreen,opacity=.8] (0,0) rectangle (2.35,4.13);
\end{scope}
%face top
\begin{scope}[canvas is yx plane at z=2.35]
  \shade[left color=mygreen,right color=green,opacity=.8] (0,0) rectangle (2.35,4.13);
\end{scope}
\draw[black] (0,0,-5.)  node {d) Swollen \& loaded};
\draw[black] (0,0,-6.5) node {(asymptotic response)};
\draw[black] (0,-2,4.)  node{$J_d=J_o\,\lambda_1\,\lambda^2$};
\end{tikzpicture}

\caption{From left to right: a) \b{dry state; b) free-swollen state;
c) fast response state under swelling and loading; d) asymptotic state under swelling and loading.} 
For a gel with  G=0.1 MPa, $R\,T/\Omega=40$ MPa,  $\chi=0.2$, under a traction $\sigma_e=0.1$ MPa, we have: $J_o=18.5$, $J_d\simeq 23$. Plots are in scale.  
\label{figF:3}}
\end{figure*}
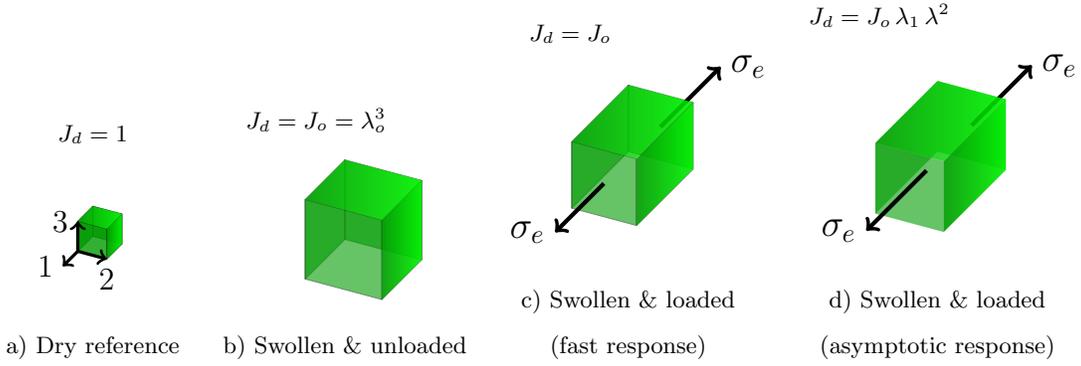
%
%%%%%%%%%%%%%%%%%%%%%%%%%%%%%%%%%%%%%%%%%%%%%%%%%%%%%%%%%%%%%
\b{We consider a dry-reference cubic gel $\Bc_d$, 
whose edges are aligned along the directions of the orthonormal basis 
$(\eb_1,\eb_2, \eb_3)$ of the three--dimensional vector space $\Vc$ (see figure \ref{figF:3}, panel a),
and its free-swollen state $\Bc_o$,  determined by the value $\mu_o$ of the solvent's bath chemical potential (see figure \ref{figF:3}, panel b).}
\b{The gel may undergo further deformations, determined by a change $\mu_e-\mu_o$ of the solvent bath's conditions, by an uniaxial boundary loads $\sigma_e$ per unit current area, and by an uniaxial step deformation $\lambda_1$.}
Both uniaxial loads and deformations induce a transversely isotropic deformation process:
\begin{equation}
\Fb=\lambda_1\eb_1\otimes\eb_1 + \lambda\check\Ib\,,\quad
\check\Ib=\Ib-\eb_1\otimes\eb_1\,,
\end{equation}
where it was assumed that loads and deformations are aligned with \b{$\eb_1$.}
The stress  shares the transversely isotropic structure of the deformation $\Fb$, and is represented as $\Tb= \sigma_1\eb_1\otimes\eb_1 + \sigma\check\Ib$. From equations \eqref{csteqn}$_1$, \eqref{mu_d}$_1$, \eqref{nH}, \eqref{pufb}$_2$, we can write the constitutive equations of $\sigma_1$ and $\sigma$  as
\begin{equation}\label{sss}
\sigma_1=\frac{G}{\lambda^2\lambda_o}\,\lambda_1 - p\,\quad\textrm{and}\,\quad
\sigma=\frac{G}{\lambda_1\lambda_o} - p\,.
\end{equation}
\b{The slow response of the gel is described by the steady solution of the  stress--diffusion problem,
under the following boundary conditions}
\[
\Tb\nb=
\begin{cases}
\sigma_e\,\eb_1			&\text{for}\quad\nb=\eb_1\\
\0		&\text{for}\quad\nb=\eb_2\,,\eb_3
\end{cases}
\quad\textrm{and}\quad
\mu=\mu_e\,
\quad
\textrm{on}\,\partial\Bc_t\,.
\]
Hence, the homogeneous solutions of the problem correspond to the  mechanical and chemical balance equations which  prescribe that
\begin{equation}\label{eqnBAL2}
\sigma_1=\sigma_e\,,\,\sigma=0\,,\quad\textrm{and}\quad
\mu=\mu_e\qquad\textrm{in}\,\, \Bc_t\,.
\end{equation}
The chemical balance  \eqref{eqnBAL2}$_3$, together with equations \eqref{csteqn}$_2$, yields the value attained by pressure $p$ at the steady state:
\begin{equation}\label{pchem}
p= \frac{1}{\Omega}(\mu_e-\mu(J\,J_o))\,. 
\end{equation}
On the other hand, the mechanical balance \eqref{eqnBAL2}$_2$, together with equation \eqref{sss}$_2$ 
\b{gives $p=G/\lambda_1\lambda_o$.} Hence, using this latter into  equations (\ref{eqnBAL2})$_1$ together with equation \eqref{sss}$_1$, and into equation \eqref{pchem}  together with equation \eqref{mud}, we get the two equations governing the steady response of the gel:
\begin{eqnarray}
&&\frac{G}{\lambda_o}\Bigl(\frac{\lambda_1}{\lambda^2} - \frac{1}{\lambda_1}\Bigr)=\sigma_e\,,\nonumber\\
&& \mu(JJ_o) + \Omega\frac{G}{\lambda_o}\frac{1}{\lambda_1} = \mu_e\,,\quad J=\lambda_1\lambda^2\,.\label{eqnani}
\end{eqnarray}
Equations \eqref{eqnani} \b{give} the stretches $\lambda_1$ and $\lambda$ which define the shape of the gel parallelepiped $\Bc_t$ under the pair $(\sigma_e,\mu_e)$; alternatively, 
\b{they} characterize the normal boundary traction $\sigma_e$, and the stretch $\lambda$ 
under the pair $(\lambda_1,\mu_e)$. 
\begin{figure}[h]
\centering
\includegraphics[scale=.35]{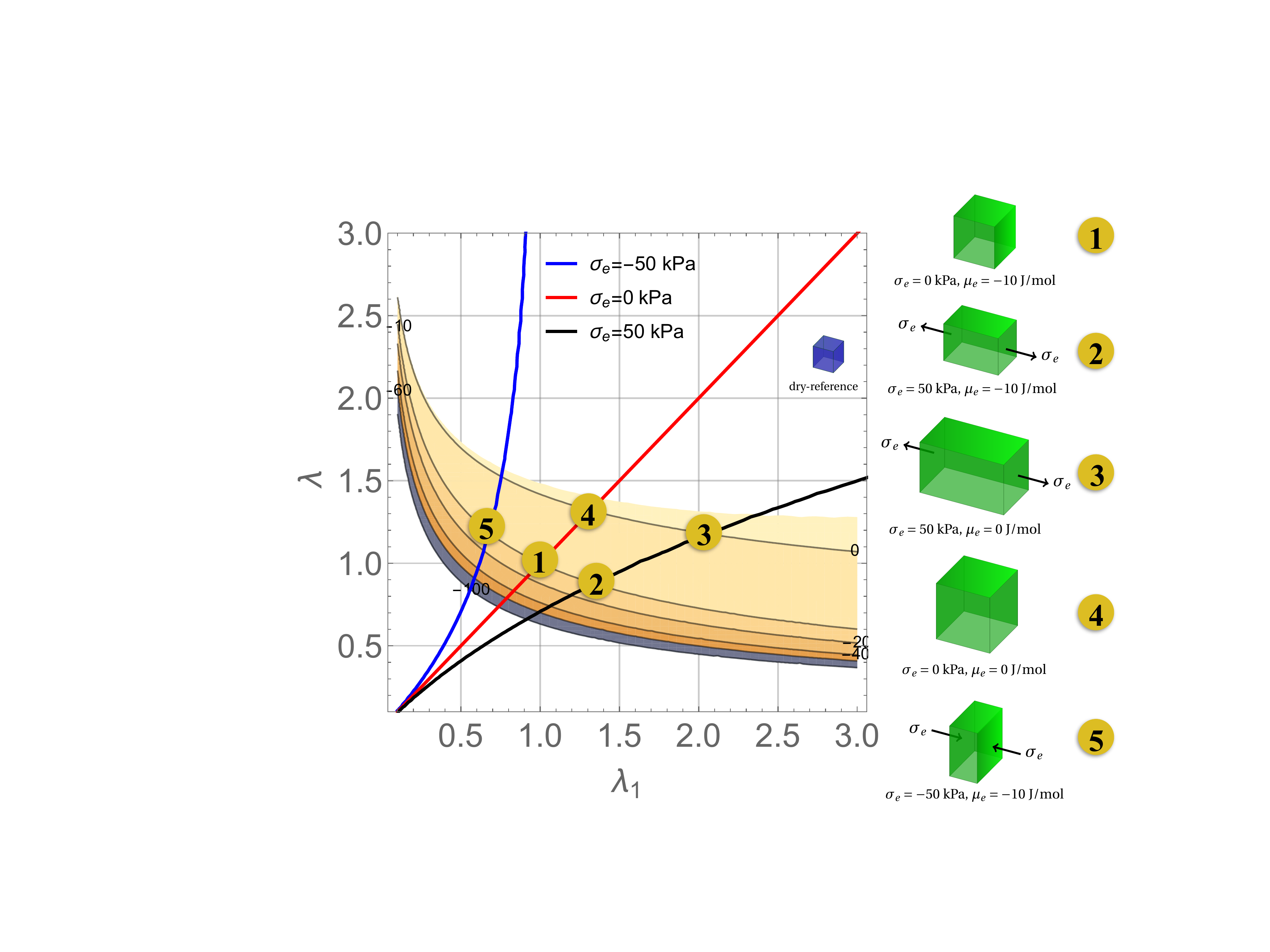}
\caption{Iso--traction lines $\sigma_1=-50$ kPa (blue line), $\sigma_1=0$ kPa (red line), $\sigma_1=50$ kPa (blue line) 
in the $\lambda_1$--$\lambda$ plane over contour lines of $\mu_e$ (from $-100$ J/mol (brown) to $0$ J/mol (light yellow).
The state of the gel is pinpointed by the intersection of iso-traction
lines with iso-potential lines: the circled numbers from $1$ to $5$ 
show five of such points on the line $\mu=-10$ J/mol (states $1$, $2$, and $5$) and on the line $\mu=0$ J/mol (states $3$ and $4$). On the right, 
\b{the corresponding shapes of the gel are shown in scale}}
\label{figF:4}
\end{figure}

\b{This} solution is shown in the \r{$\lambda_1$--$\lambda$} plane in figure \ref{figF:4}, where the brown--to--orange colours identify different values of  $\mu_e$ (from $-100$ to $0$ J/mol), whereas $\mu_o$ has been fixed as $\mu_o=-10$ J/mol. The red line is the iso--$\sigma_1$ (or equivalently, due to equation (\ref{eqnBAL2}), iso--$\sigma_e$) line corresponding to $\sigma_1=0$ kPa; the corresponding $\lambda_1=\lambda$ values identify the size of the free--swelling states, at  different values of $\mu_e$. We can move from  state $1$ ($\lambda_1=\lambda=1$) to state $2$ along the iso--$\mu_e$ line $\mu_e=-10$ J/mol, acting upon the gel with a traction $\sigma_e=50$ kPa; then, from state $2$ to state $3$, keeping the traction fixed and increasing the chemical potential to a new value $\mu_e=0$ J/mol; and from state $3$ to state $4$ along the new iso--$\mu_e$ line $\mu_e=0$ J/mol removing the traction; and, at the end,  go back to state $1$ only changing $\mu_e$ to the old value $\mu_e=-10$J/mol. 
The corresponding \b{shapes of the cubic gel are shown, in scale, in the lateral panel in figure \ref{figF:3}.}

Figure \ref{figF:4} also allows to  evaluate the traction $\sigma_e$ corresponding to an imposed deformation $\lambda_1$ (which might represent the value prescribed by a boundary constraint), for different values of $\mu_e$. As an example, for $\lambda_1=0.68$ we get $\sigma_e=-50$ kPa at $\mu_e=-10$ J/mol; it means that the freely swollen cube $\Bc_o$ with sides's length equal to $\lambda_o$, once constrained to reduce the length of the side aligned with $\eb_1$ to $\lambda_1\lambda_o= 0.68\lambda_o$ would exert a traction equal to $-\sigma_e=50$ kPa on the constraint.
\begin{figure}[h]
\centering
\includegraphics[scale=.6]{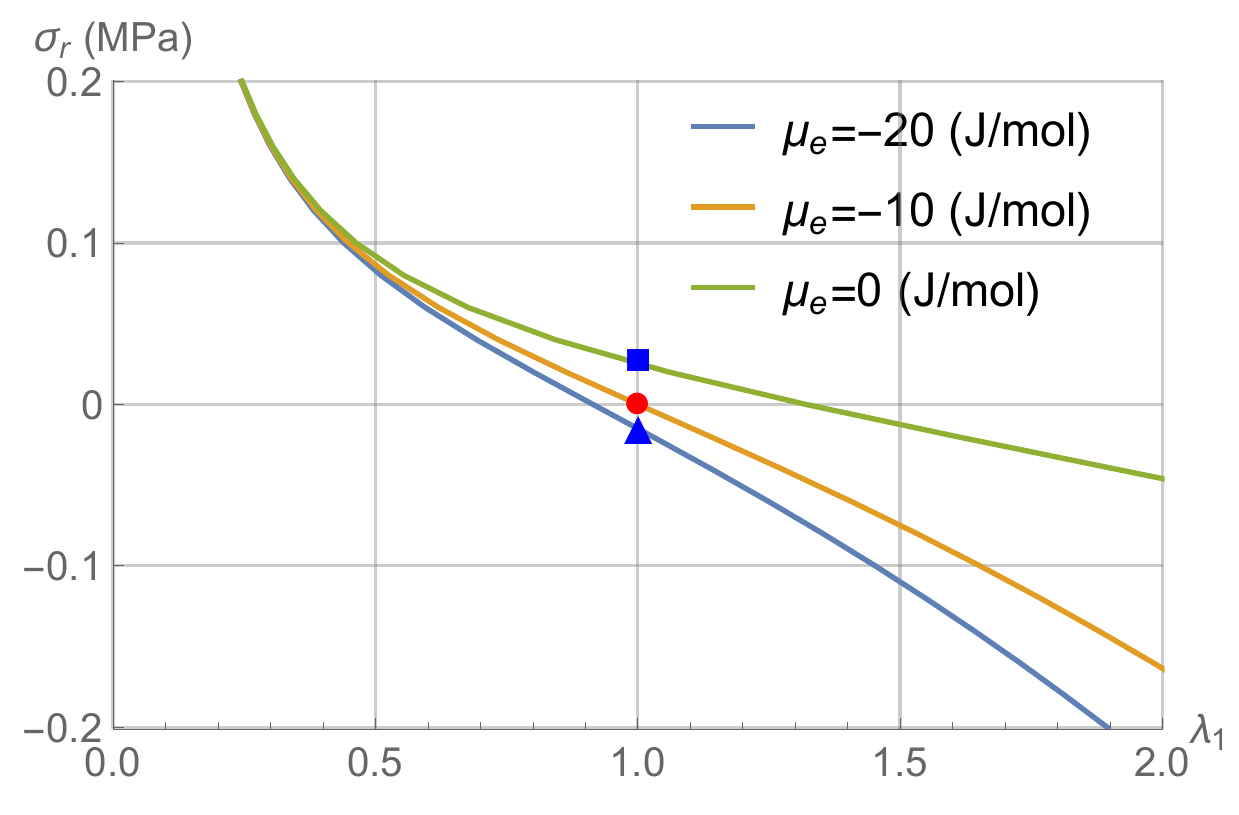}
\caption{Stress--stroke curves $\sigma_r=- \sigma_e(\lambda_1)$ at different values of $\mu_e$ when $\mu_o=-10$ J/mol and $\lambda_o=2$, being $G=0.1$ MPa.}
\label{figF:5}
\end{figure}

A direct \b{visualization} of the blocking force $\sigma_r=-\sigma_e$, that is, 
\b{of the traction exerted on the constraints hampering the deformation of the swollen state $\Bc_o$ 
along $\eb_1$, is given in  figure  \ref{figF:5}: the blue square (triangle) on the green (blue) line identifies the traction acting on the constraint that maintains $\lambda_1=1$.} It is worth noting that, being the gel isotropic, a completely equivalent situation would correspond to a constraint which hamper the full deformation along the direction spanned by $\eb_\alpha$, with $\alpha=2,3$.
Our results are very similar to the ones in Ref.\citenum{Illeperuma2013}, where the same problem was discussed for isotropic temperature--sensitive hydrogels both from an experimental and theoretical point of view, 
through the so--called ideal elastomeric gel model. 

For large swelling--induced deformations, \textit{i.e.} $1/J_o\to 0$ and  $1/J\to 0$, equation \eqref{pchem} can be approximated as
\begin{equation}
p\simeq - \frac{RT}{\Omega}(\chi-1/2)\frac{1}{J_o^2J^2}\simeq\frac{G}{\lambda_o}\frac{1}{\lambda_1^2\lambda^4}\,,
\end{equation}
where we used equation \eqref{eqcap} and set $\mu_o=\mu_e=0$. 
\b{With this, being $\sigma=0$,} we get the relationship which links the steady values  $\lambda_1$ and $\lambda$ (experimentally discussed in Ref.\citenum{Urayama:2012}), as well as the asymptotic relationship \b{valid} at the steady state between  $\sigma_1$  and $\lambda_1$:
\begin{equation}\label{steadyeqn}
\lambda=\lambda_1^{-1/4}\quad\textrm{and}\quad \sigma_1=\frac{G}{\lambda_o}(\lambda_1^{3/2}-\lambda_1^{-1})\,.
\end{equation}
\subsection{Fast response of isotropic gels}
Figure \ref{figF:4} describes the \b{asymptotic} state under uniaxial step traction or deformation. 
\b{It is also of interest to determine the mechanical state just after traction (or deformation) is applied, and before diffusion starts; during such a transient, the gel behaves as an elastic, and incompressible solid:}
%Precisely, fixed the traction $\sigma_e$ and known $\mu_e$, 
%it allows to determine the steady values attained by the 
%stretches $\lambda_1$ and $\lambda$; alternatively, 
%fixed the extension $\lambda_1$ and known $\mu_e$, it allows to determine the steady values 
%attained by the stretch $\lambda$ and the erogated traction $\sigma_e$. 
%However, just after traction (or deformation) is applied and before diffusion starts, 
%gel behaves as an elastic incompressible solid:
%
\begin{equation}\label{fastiso}
J=1\quad\textrm{and}\quad\lambda=\lambda_{1}^{-1/2}\,;
\end{equation}
moreover, its response is different. Indeed, from being $\sigma=0$, we get the relationship holding between the before--diffusion--starts values $\sigma_{1f}$ of the stress and $\lambda_1$ :
\begin{equation}\label{bdseqn}
\sigma_{1f} = \frac{G}{\lambda_o}\Bigl(\lambda_{1}^2 - \frac{1}{\lambda_{1}}\Bigr)\,,
\end{equation}
where the equation \eqref{fastiso}$_2$ has been taken into account.

\begin{figure}[h]\centering
\begin{tikzpicture}[scale=1,baseline]
	\begin{axis}	[yscale=0.6,xscale=1,grid=major,minor tick num=3,
	             xmin=0.1,xmax=3,
	             ymin=-10,ymax=30,
	             xlabel={$\lambda_1$},
	             ylabel={$\sigma_1\,\lambda_o/G\,, \sigma_{1f}\,\lambda_o/G$},
	             legend style={ at={(.515,1.23)}, anchor=east}]
   \addplot[red,very thick,domain=0.:3,smooth,samples=40] {(x^(3/2)-1/x)};
   \addlegendentry{Asympt. (iso)}
   \addplot[blue,very thick,domain=0.:3,smooth,samples=40]   {(x^2-1/x)};
   \addlegendentry{Fast (iso)}
   \addplot[red,dashed,very thick,domain=0.:3,smooth,samples=40] 
       {((1 + 2*1*(2^2 + x^2 - 1))/(1 + 2*1*(2^2 - 1))*x^(3/2)-1/x)};
   \addlegendentry{Asympt. (aniso)}
   \addplot[blue,dashed,very thick,domain=0.:3,smooth,samples=40] 
       {((1 + 2*1*(2^2 + x^2 - 1))/(1 + 2*1*(2^2 - 1))*x^2-1/x)};
   \addlegendentry{Fast (aniso)}
   \end{axis}
\end{tikzpicture}

\caption{ \label{figF:6} The swelling--induced stress reduction  is measured by the difference between the blue solid line (equation \eqref{bdseqn}) and the red solid line (equation \eqref{steadyeqn}) at the same value of the imposed deformation $\lambda_1$. That difference represents material relaxation due to diffusion. The corresponding dashed lines measure the swelling--induced stress reduction in presence of fibers ($\gamma=1$ and $\lambda_{o\parallel}=2$) as difference between the blue dashed line (equation \eqref{fastani}) and the red dashed line (equation \eqref{slows}) at the same value of the imposed deformation $\lambda_1=\lambda_\parallel$.}
\end{figure}
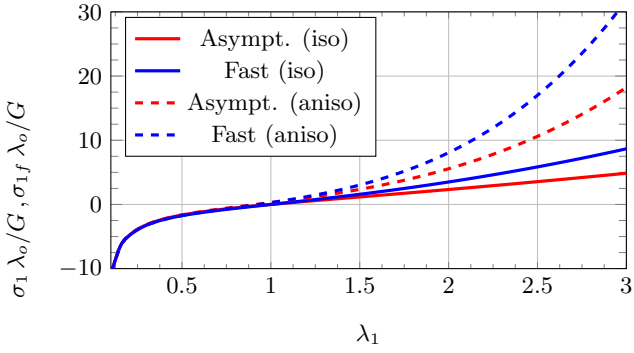

A comparison between equations \eqref{steadyeqn} and \eqref{bdseqn} allows to identify both the force relaxation due to a step deformation and the creep due \b{to diffusion in response to a step load, a phenomenon which is
different from the one} characteristic of the viscoelastic response of solids, as discussed in Ref.\citenum{Urayama:2012}. \b{Figure \ref{figF:6} shows the stress versus the stretch $\lambda_1=\lambda_{1f}$ for
both the fast and the asymptotic response.}
%%
%\begin{figure}[h]\centering
%\begin{tikzpicture}[scale=1,baseline]
%\begin{axis}
%[yscale=0.6,xscale=1,grid=major,minor tick num=3,
%            xmin=0.1,xmax=3,
%            ymin=-10,ymax=10,
%            xlabel={$\lambda_1$},
%            ylabel={$\sigma_1\,\lambda_o/G\,, \sigma_{1f}\,\lambda_o/G$},
%            legend style={ at={(1.,.5)}, anchor=east}]
%% 
%   \addplot[red,very thick,domain=0.:3,smooth,samples=40] {(x^(3/2)-1/x)};
%   \addlegendentry{Asymptotic response}
%   %
%   \addplot[blue,very thick,domain=0.:3,smooth,samples=40]   {(x^2-1/x)};
%   \addlegendentry{Fast response}
% %  \addplot[black,very thick,domain=1.01:1.1,samples=20] {1};
%%    \addplot[no markers] coordinates {(0,0.772) (0.54,0.772)};  
%   \end{axis}
%\end{tikzpicture}
%%
%\caption{The swelling--induced stress reduction  is measured by the difference between the blue line (equation \eqref{bdseqn}) and the red line (equation \eqref{steadyeqn}) at the same value of the imposed deformation $\lambda_1$. That difference represents material relaxation due to diffusion. \label{figF:6}}
%\end{figure}
%%
%
%%%%%%%%%%%%%%%%%%%%%%%%%%%%%%%%%%%
\section{Anisotropic gels under uniaxial traction and extension}
\begin{figure}[h]
\centering
\includegraphics[scale=.4]{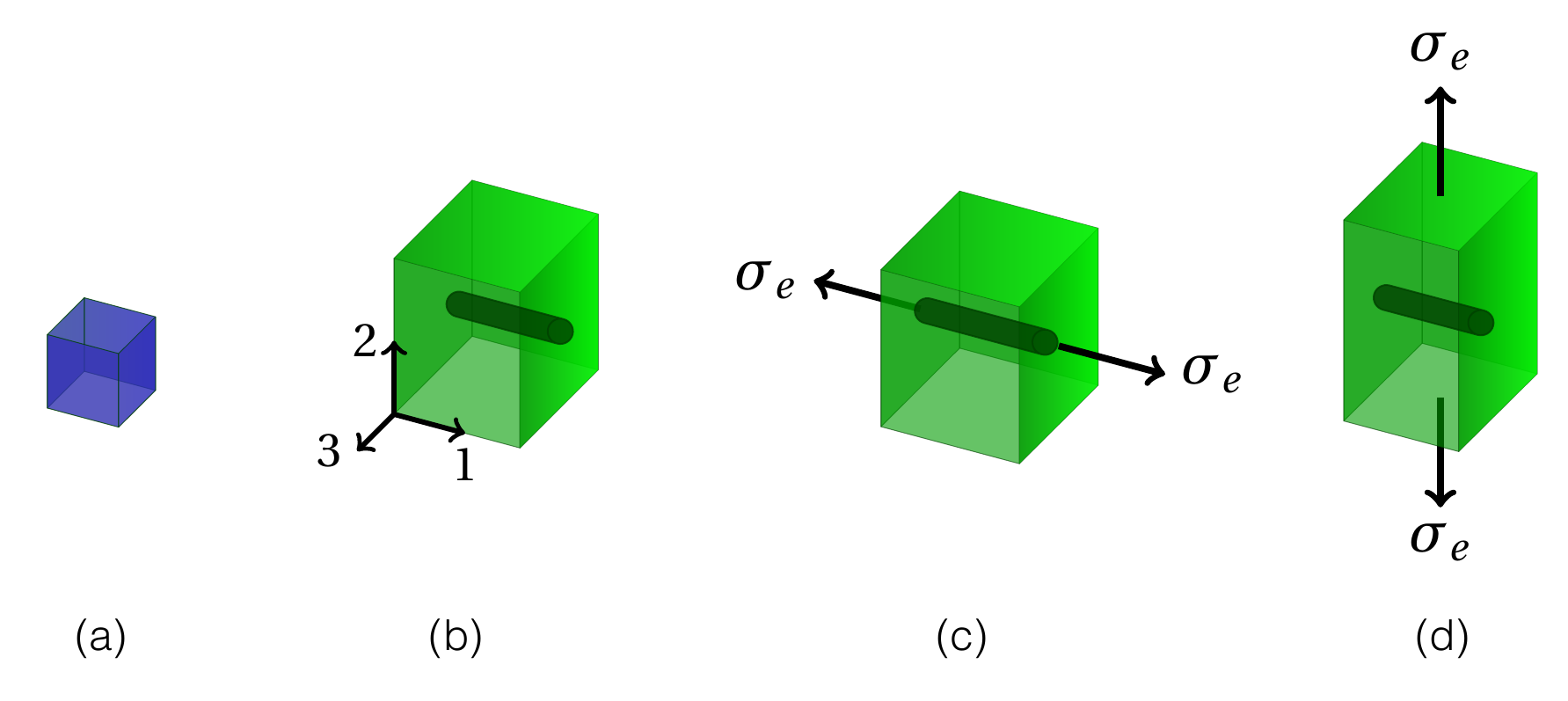}
\caption{\b{From left to right: (a) dry state; (b) free--swollen state: $\sigma_e=0$, and $\mu_e=-10$ J/mol; 
(c) swollen with traction along the fiber: $\sigma_e=50$ kPa, and $\mu_e=-10$ J/mol; 
(d) swollen with traction orthogonal to the fiber: $\sigma_e=50$ kPa, and $\mu_e=-10$ J/mol.} }
\label{fig:5b}
\end{figure}

For anisotropic gels, we proposed in Ref.\citenum{SM2015} an extension of the classical Flory--Rehner \b{model, 
where the elastic term $\psi_e$ in the free energy $\psi$ has the following form:}
\begin{equation}\label{aniFR}
\psi_e(\Fb_d)= \frac{G}{2}(\Fb_d\cdot\Fb_d-3) + \frac{1}{2}G\,\gamma\,(\Fb_d\,\eb\cdot\Fb_d\,\eb-1)^2\,,
\end{equation}
with $\gamma$ \b{is a stiffening parameter, and the unit vector $\eb$ describes the fibers direction. Behind the representation \eqref{aniFR}, there is the idea to describe the effect of the presence of reinforcements (fibers) into the gel, which hamper the swelling along their direction.}  
Within that context, the dry--reference stress $\Sb_d$ is represented by equation \eqref{csteqn}$_1$ with
\begin{eqnarray}\label{csteqnani}
\Sb_d(\Fb_d)  &=& \b{G\,\Fb_d + 2\,G\,\gamma\,(\Cb_d\cdot\Eb-1)\Fb_d\,\Eb\,,}
\end{eqnarray}
being $\Cb_d=\Fb_d^T\,\Fb_d$ the right Cauchy-Green strain, 
and $\Eb=\eb\otimes\eb$ the direction of the fiber $\eb$. 
We are interested in homogeneous solutions of the swelling problem which realize the following 
\b{triaxial} deformation $\Fb_d$:
\begin{equation}\label{ciccio}
\Fb_d = \lambda_{di}\Eb_i\,,\quad \Eb_i= \eb_i\otimes\eb_i\,,\,i=1,2,3\,,
\end{equation}
compatible with fibres aligned with $\eb_1$ or $\eb_2$ or $\eb_3$.
As $\Bb_d=\Cb_d=\lambda_{di}^2\,\Eb_i$, the (Cauchy) stress $\Tb$ admits the following representation
\begin{equation}\label{scauchy}
\Tb = \frac{1}{J_d}\,G\,\Bigl(\Cb_d 
     + 2\,\gamma\,(\Cb_d\cdot\Eb-1)\Fb_d\,\Eb\,\Fb_d^T\Bigr)-p\,\Ib\,.
\end{equation}
\b{Given a plane in the dry-reference configuration having unit normal $\mb$, 
the image under $\Fb_d$ of that plane will have a normal $\nb$ represented by}
\begin{equation}
\nb=\frac{\Fb_d^\star\,\mb}{\vert\Fb_d^\star\,\mb\vert}
   =\frac{\lambda_i^{-1}\,\Eb_i\,\mb}{\vert\lambda_i^{-1}\,\Eb_i\,\mb\vert}\,;
\end{equation}
\b{Denoting with $\tb$ a unit vector orthogonal to $\nb$, 
the normal} and tangential stress component of $\Tb$ with respect to $\nb$ and $\tb$ 
are $\sigma_\nb=\Tb\,\nb\cdot\nb$ and $\tau_\nb=\Tb\,\nb\cdot\tb$, and are \b{given by:}
\begin{eqnarray}
\sigma_\nb &=&\frac{G}{J_d}\,\lambda_{di}^2\,(\nb\cdot\eb_i)^2 -p\label{tractions}\\
&+&2 \,\frac{G\,\gamma}{J_d}\,(\lambda_{di}^2\,(\eb\cdot\eb_i)^2-1)\,
      (\lambda_{di}\,(\nb\cdot\eb_i)\,(\eb\cdot\eb_i))^2\,,\nonumber\\
\tau_\nb &=&\frac{G}{J_d}\,\lambda_{di}^2\,(\nb\cdot\eb_i)\,(\tb\cdot\eb_i)
          + 2\,\frac{G\,\gamma}{J_d}\,(\lambda_{di}^2\,(\eb\cdot\eb_i)^2-1)\nonumber\\
&\cdot&(\lambda_{di}\,(\nb\cdot\eb_i)\,(\eb\cdot\eb_i))\,
       (\lambda_{di}\,(\tb\cdot\eb_i)\,(\eb\cdot\eb_i))\,.\nonumber
\end{eqnarray}
\b{For anisotropic gels, free--swelling states yield changes in both size and shape.\cite{JAP2015}
As for isotropic gel, balance laws prescribe $\Sb_d=\0$, and $\mu_e=\mu_o$; 
the free-swollen state $\Bc_o$, deformed by $\Fb_d=\Fb_o=\lambda_{o\parallel}\,\Eb+\lambda_{o\perp}\,\hat\Ib$
with respect to $\Bc_d$,  is completely defined by the value $\mu_o$ of the bath's chemical potential. 
Let us only note that, for $\mu_o=0$, the relationship between $\lambda_{o\parallel}$ and $\lambda_{o\perp}$ prescribes that}
\begin{equation}\label{fsani1}
\lambda_{o\perp}^2=\lambda_{o\parallel}^2(1+2\gamma(\lambda_{o\parallel}^2-1))\,,
\end{equation}
and, for large deformation ($1/J_o\to 0$), it holds
\begin{equation}\label{cap1}
\frac{RT}{\Omega}(\chi-1/2)\frac{1}{J_o^2}=-\frac{G}{\lambda_{o\parallel}}\,,\quad
J_o=\lambda_\parallel\lambda_\perp^2\,.
\end{equation}
In the following, we  present and discuss the slow and fast responses of anisotropic  gels  to imposed uniaxial tractions and/or deformations, with reference to an unit cube $\Bc_d$ at dry state, 
\b{having fibers} aligned along the direction $\eb=\eb_1$ (panel (a) in figure \ref{fig:5b}), 
which \b{realizes a free--swollen state $\Bc_o$ with a bath's chemical potential equal to $\mu_o$} (panel (b) in figure \ref{fig:5b}). 
\b{The free-swollen gel may experience a further deformation determined by a change of the bath's potential, 
by the action of uniaxial boundary loads $\sigma_e$ per unit current area, or by a uniaxial step extension.}
%
%As the gel body is anisotropic, applying normal boundary tractions as well as imposing deformations 
% along different directions deliver different shapes. 
%

We shall examine two cases: (A) corresponds to normal boundary tractions on the faces of unit normal $\pm\eb_1$ or, equivalently, to impose a deformation $\lambda_1$ (Section \ref{parallel}); 
(B) corresponds to normal boundary tractions on the faces of unit normal $\pm\eb_2$ or, equivalently, 
to impose a \b{stretch of intensity $\lambda_2$ along $\eb_2$} (Section \ref{transverse}). 
\par
\subsection{Parallel--to--the--fibers normal loads}
\label{parallel}
\b{The deformation is transversely isotropic; thus, $\lambda_{d1}=\lambda_{d\parallel}$,
 and $\lambda_{d2}=\lambda_{d3}=\lambda_{d\perp}$, being $\lambda_{d\parallel}$ and $\lambda_{d\perp}$  
 the linear swelling ratios along the fiber direction $\Eb=\Eb_1$ and in the orthogonal plane $\check\Ib=\Ib-\Eb=\Eb_2+\Eb_3$, respectively:}
\begin{equation}
\Fb_d=\lambda_{d\parallel}\Eb+\lambda_{d\perp}\check\Ib\,.
\end{equation} 
\begin{figure}[h]
\centering
\includegraphics[scale=.45]{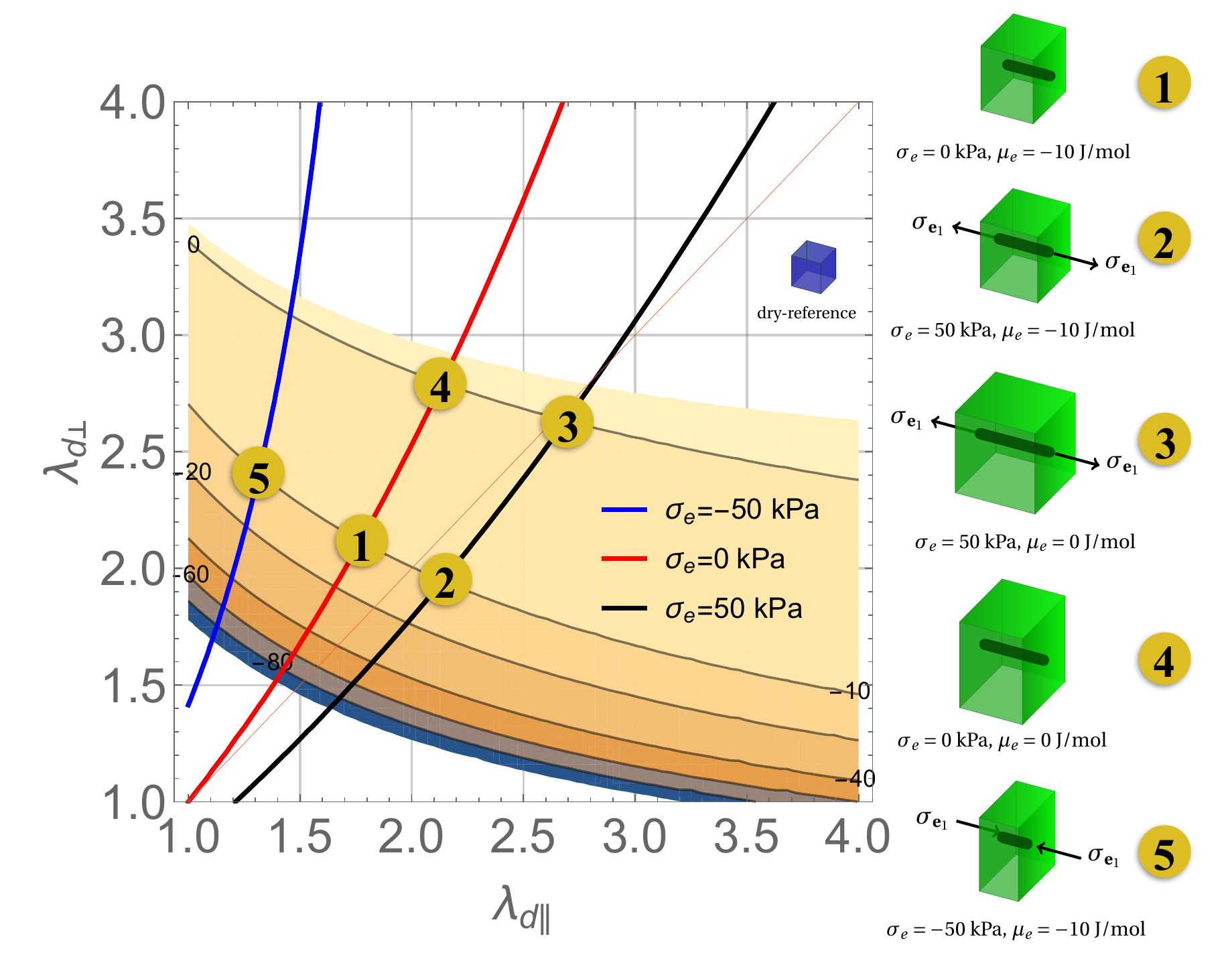}
\caption{Iso--traction lines $\sigma_e=-50$ kPa (blue line), $\sigma_e=0$ kPa (red line), $\sigma_e=50$ kPa (black line) in the $\lambda_{d1}$--$\lambda_{d2}$ plane, over contour lines of $\mu_e$ (from $-100$ J/mol (brown) to $0$ J/mol (light yellow).
\b{The state of the gel is pinpointed by the intersection of iso-traction
lines with iso-potential lines: the circled numbers from $1$ to $5$ 
show five of such points on the line $\mu=-10$ J/mol (states $1$, $2$, and $5$) and on the line 
$\mu=0$ J/mol (states $3$ and $4$). On the right, the corresponding shapes of the gel are shown in scale}.}
\label{fig:5d}
\end{figure}
\par\noindent
We can set  the stress--diffusion problem in the dry state $\Bc_d$ and look for homogeneous steady solutions which are driven  by the boundary conditions
\[
\Sb_d\mb=
\begin{cases}
s_e\eb_1			&\text{for}\,\,\mb=\eb_1\,,\\
\0		&\text{for}\,\,\mb=\eb_2\,,\eb_3\,,
\end{cases}
\quad\textrm{and}\quad
\mu=\mu_e\,,
\]
on $\partial\Bc_d$, being $s_e=\lambda_{d\perp}^2\sigma_e$ the uniaxial and normal boundary load per unit dry area, and  $\sigma_e$ the corresponding load per unit of current area (see panel (c) in figure \ref{fig:5b}). Hence,  mechanical and chemical balances prescribe
\begin{equation}\label{eqnBAL3}
s_{d\parallel}=s_e\,,\,s_{d\perp}=0\,,\quad\textrm{and}\quad
\mu=\mu_e\,\,\textrm{in}\,\, \Bc_d\,,
\end{equation}
being $\Sb_d=s_{d\parallel}\Eb + s_{d\perp}\check\Ib$ the representation of the dry--reference stress.  
The anisotropy of the gel is of elastic nature, and does not change the constitutive structure of the chemical potential; hence, the chemical balance  \eqref{eqnBAL3}$_3$, together with equations \eqref{csteqn}$_2$, yields the osmotic pressure field in the form given by the equation \eqref{pchem}. Once used this last  equation  into  \eqref{csteqnani} and \eqref{csteqn}$_1$,  the mechanical balance  \eqref{eqnBAL3}$_2$ yields
\begin{equation}\label{case1}
\mu(J_d) + \frac{\Omega\,G}{\lambda_{d\|}}=\mu_e\,,\quad J_d=\lambda_{d\|}\,\lambda_{d\perp}^2\,;
\end{equation}
this last relation characterizes $\lambda_{d\perp}$ in terms of $\mu_e$ and $\lambda_{d\|}$: $\lambda_{d\perp}=\lambda_{d\perp}(\lambda_{d\|},\mu_e)$. Equations \eqref{pchem} and \eqref{csteqnani} allow to evaluate the stress component $s_{d\parallel}$. Being  
$s_{d\parallel}=s_e=\lambda_{d\perp}^2\sigma_e$, we get 
\begin{equation}\label{traction1}
\sigma_e= \frac{G\,\lambda_{d\|}}{\lambda_{d\perp}^2}
\Bigl(1+2\,\gamma\,(\lambda_{d\|}^2-1)\Bigr)-\frac{G}{\lambda_{d\|}}\,.
\end{equation}
Equation \eqref{traction1} also delivers $\sigma_{\eb_1}$, being this latter equal to $\sigma_e$; 
from equation \eqref{tractions}, we get as expected $\tau_{\eb_1}=0$, \textit{i.e.} no tangential tractions are exerted on the face of unit normal $\eb_1$.

Equations  \eqref{case1} and \eqref{traction1} determine $\lambda_{d\|}$ and $\lambda_{d\perp}$ in terms of the pair $(\sigma_e,\mu_e)$, when these latter are the control parameters of the deformation process or, equivalently, allow to evaluate $\lambda_{d\perp}$ and $\sigma_e$ in terms of the pair $(\lambda_\parallel,\mu_e)$ when a deformation $\lambda_\parallel=\lambda_{d\parallel}/\lambda_{o\parallel}$ is imposed on the free--swollen state $\Bc_o$. 
 
The solution is shown in the $\lambda_{d\|}-\lambda_{d\perp}$ plane in figure \ref{fig:5d} for $G=0.1$ MPa and $\gamma=0.1$, where we \b{used the same color code as in figure \ref{figF:4} for $\mu_e$, 
ranging from $-100$ J/mol (brown) to $0$ J/mol (light yellow).}
The red line is the iso--$\sigma_e$  line corresponding to $\sigma_e=0$, which is no longer along the bisectrix of the plane due to the anisotropic swelling: $\lambda_{d\|}$ is always smaller than $\lambda_{d\perp}$. 

The corresponding $\lambda_{d\|}$ and $\lambda_{d\perp}$ values identify the size of the free--swelling states, at  different values of $\mu_e$. 
We can move from  state $1$ ($\lambda_{d\parallel}= 1.77$ and $\lambda_{d\perp}=2.12$) to state $2$ along the iso-potential line $\mu_e=-10$ J/mol, acting upon the gel with a traction $\sigma_e=50$ kPa; then, from $2$ to $3$, 
along the iso-traction line $\sigma_e=50$ kPa, by increasing the chemical potential to the new value $\mu_e=0$ J/mol; 
from $3$ to $4$ along the new iso-potential line $\mu_e=0$ J/mol, and removing the traction; then, eventually,  
go back to state $1$ by decreasing $\mu_e$ to the initial value $\mu_e=-10$ J/mol. 
The corresponding actual shapes realized by the gel cube are shown, \b{in scale,} in the lateral panel in figure \ref{fig:5d}.

%\subsubsection{Asymptotic and fast response}
%
We may compare asymptotic and fast response of isotropic gels under uniaxial traction with the corresponding responses of anisotropci gels under uniaxial traction aligned with fiber direction. For large swelling--induced deformations, \textit{i.e.} $1/J_o\to 0$ and  $1/J\to 0$, equation \eqref{pchem} can be approximated as
\begin{equation}\label{asym}
p\simeq - \frac{RT}{\Omega}(\chi-1/2)\frac{1}{J_o^2J^2}\simeq\frac{G}{\lambda_{o\parallel}}\frac{1}{J^2}\,,
\end{equation}
where we set $\mu_o=\mu_e=0$. Under the asymptotic approximation determined by the equation \eqref{asym},  
mechanical balance \eqref{eqnBAL3}$_{1,2}$ \b{yields} 
$\lambda_\perp=\lambda_\parallel^{-1/4}$ and  $\sigma_{\eb_1}=\sigma_e$ as
\begin{equation}\label{slows}
\sigma_{\eb_1}=\frac{G}{\lambda_{o\parallel}}\,\Bigl(\alpha(\lambda_{\parallel},\gamma)\,\lambda_{\parallel}^{3/2} 
              - \lambda_\parallel^{-1}\Bigr)\,,
\end{equation}
with 
\begin{equation}\label{alfa}
\alpha(\lambda_{\parallel},\gamma)=\frac{1+2\,\gamma\,(\lambda_{o\parallel}^2\,\lambda_{\parallel}^2-1)}
                                  {1+2\,\gamma\,(\lambda_{o\parallel}^2-1)}\,.
\end{equation}
The relationship between the two stretches is the same as in the isotropic case, and says that Poisson modulus at the steady state is $1/4$; on the other hand, as $\alpha(\lambda_\parallel,0)=1$,  for $\gamma\to 0$ equation \eqref{slows} delivers equation \eqref{steadyeqn}$_2$ and the isotropic case is recovered.

 The fast response to the deformation $\lambda_{\parallel}$ is driven by the anisotropic elastic nature of the network. Before diffusion starts, we have $J=\lambda_\parallel\lambda_\perp^2=1$; and equation \eqref{eqnBAL3}$_2$ determines the pressure field $p$ and 
\begin{equation}\label{fastani}
\sigma_{f\eb_1}=\frac{G}{\lambda_{o\parallel}}\Bigl(\alpha(\lambda_\parallel,\gamma)\,\lambda_{\parallel}^2 
               - \lambda_\parallel^{-1}\Bigr)\,.
\end{equation}
The difference between the two stresses \b{is a measure of} the swelling--induced relaxation in the gel. Figure \ref{figF:6} shows the differences in stress relaxation due to the anisotropy when $\gamma=1$. It is worth noting that anisotropy enhances stress relaxation when the imposed uniaxial deformation is aligned with fiber direction.
\subsection{Transverse--to--the--fibers normal loads}
\label{transverse}
In this case, fibres and tractions are not aligned and the deformation $\Fb_d$ maintains the 
\b{triaxial anisotropic structure given by (\ref{ciccio}).} 
We look for homogeneous solutions of the stress--diffusion problem \b{posed on the dry configuration $\Bc_d$, 
under the following boundary conditions:}
\[
\Sb_d\,\mb=
\begin{cases}
\0		    	&\text{for}\quad\mb=\eb_1\,,\eb_3\,,\\
s_e\,\eb_2		&\text{for}\quad\mb=\eb_2\,,
\end{cases}
\quad\textrm{and}\quad
\mu=\mu_e\,
\]
on $\partial\Bc_d$. Hence,  mechanical and chemical balances prescribe
\begin{equation}\label{eqnBAL4}
s_{d1}=s_{d3}=0\,,\,s_{d2}=s_e\,,\quad\textrm{and}\quad
\mu=\mu_e\,\,\textrm{in}\,\, \Bc_d\,,
\end{equation}
being $\Sb_d=s_{di}\eb_i\otimes\eb_i$, $(i=1,2,3)$, the appropriate representation of the dry--reference stress.  
The anisotropy of the gel is of elastic nature, and does not change the constitutive \b{formula} of the chemical potential; hence, the chemical balance  \eqref{eqnBAL4}$_3$, together with equations \eqref{csteqn}$_2$, 
yield the osmotic pressure field in the form given by the equation \eqref{pchem}. 
Once used this last  equation  into   \eqref{csteqn}$_1$ and \eqref{csteqnani},  the mechanical balance  \eqref{eqnBAL4}$_2$ yields
\begin{equation}\label{case2}
\mu(J_d) + \frac{\Omega\,G}{\lambda_{d1}\,\lambda_{d2}}\,\lambda_{d3}=\mu_e\,,
\quad 
J_d=\lambda_{d1}\,\lambda_{d2}\,\lambda_{d3}\,.
\end{equation}
On the other hand, equation  \eqref{eqnBAL4}$_{1}$  delivers
\begin{equation}\label{num2}
\lambda_{d3}^2= (1+2\,\gamma\,(\lambda_{d1}^2-1))\,\lambda_{d1}^2\,,  
\end{equation}
\b{that is, yields the relation $\lambda_{d3}=\lambda_{d3}(\lambda_{d1})$.}
\b{Equations \eqref{case2} and \eqref{num2} concoct the representation} of  $\lambda_{d1}$ in terms of $\mu_e$ and $\lambda_{d2}$: $\lambda_{d1}=\lambda_{d1}(\lambda_{d2},\mu_e)$. 
Equations \eqref{pchem} and \eqref{csteqnani} allow to evaluate the stress component $s_{d2}$. 
Being  $s_{d2}=s_e=\lambda_{d1}\,\lambda_{d3}\,\sigma_e$, we get 
\begin{equation}\label{traction2}
\sigma_e = \frac{G}{\lambda_{d1}}\Bigl(\frac{\lambda_{d2}}{\lambda_{d3}}-\frac{\lambda_{d3}}{\lambda_{d2}}\Bigr)\,,
\end{equation}
with $\lambda_{d3}=\lambda_{d3}(\lambda_{d1})$ and $\lambda_{d1}=\lambda_{d1}(\lambda_{d2},\mu_e)$.

Equation \eqref{traction2} also delivers $\sigma_{\eb_2}$, \b{as $\sigma_{\eb_2}=\sigma_e$; from equation \eqref{tractions}$_2$, we get as expected $\tau_{\eb_2}=0$} \textit{i.e.} no tangential tractions are exerted on the face of unit normal $\eb_2$ under this deformation process.
\subsection{Blocking forces}
The characterisation of stroke curves in anisotropic gel actuator is especially interesting as both normal and tangential blocking forces can arise, depending on the anisotropic structure of the gel.  
\begin{figure}[h]
\centering
\includegraphics[scale=.45]{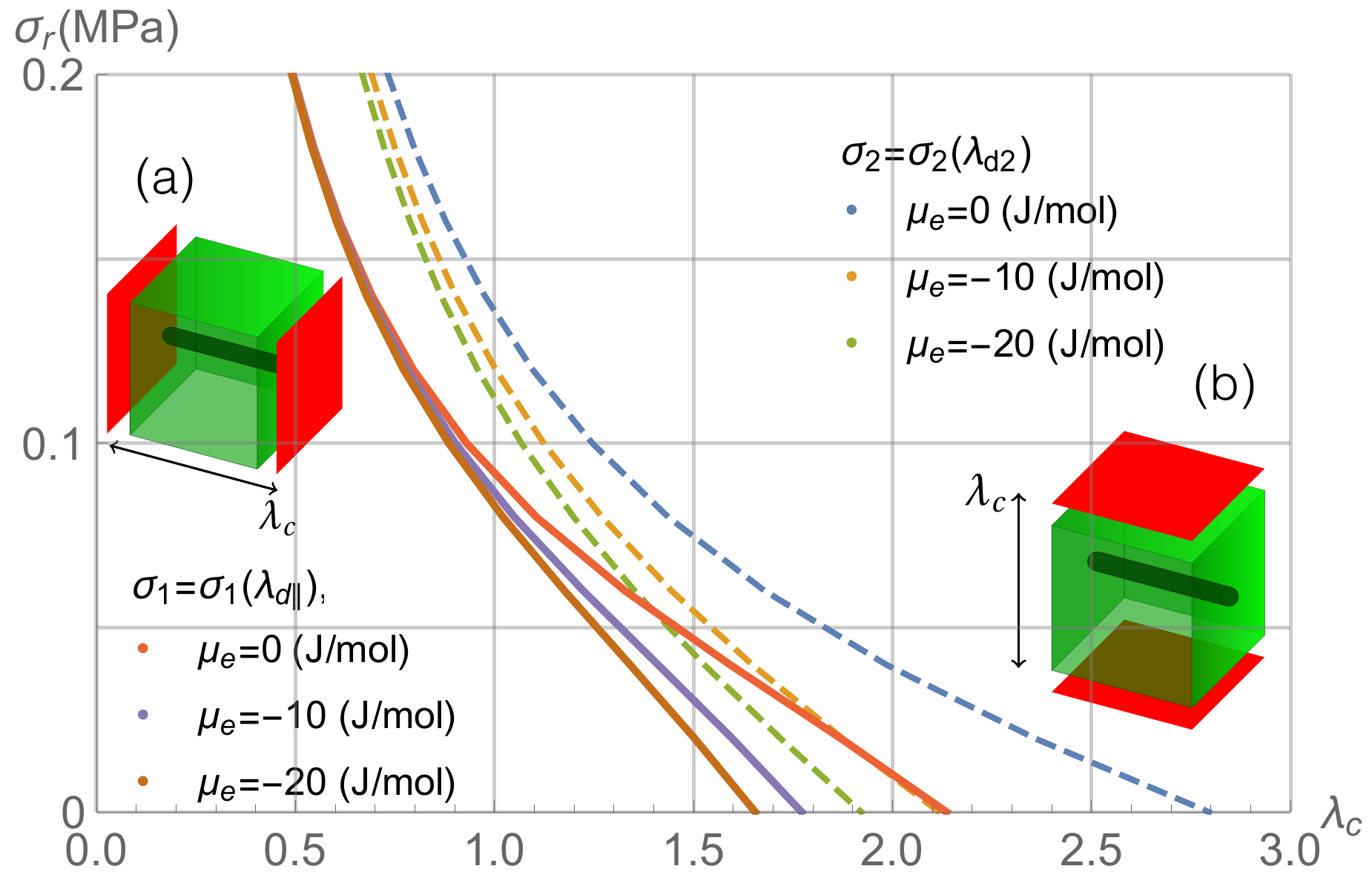}
\caption{\b{Stroke curves $\sigma_r(\lambda_c)$ for different values of $\mu_e$: 
case a) $-\sigma_1(\lambda_{d\|})$ (solid lines); case b) $-\sigma_2(\lambda_{d2})$ (dashed lines).
The insets shows the constraints (red surfaces) acting on the fibered gel cube in the two cases (a) (left) and (b) (right).}}
\label{fig:6}
\end{figure}
%

%\noindent 
In the case  parallel--to--the--fibers normal blocking forces, and with reference to Section \ref{parallel},  $\sigma_e$ may be viewed as the boundary traction exerted on the body by constraints hampering deformation along $\eb_1$ (see inset (a) in figure \ref{fig:6}). In this case,  figure \ref{fig:6} \b{shows}, for different values of $\mu_e$, the  traction corresponding to the constraints which  prescribes \b{a stretch intensity}  $\lambda_{d\parallel}=\lambda_c$. Precisely, \b{ by using equation \eqref{case1} to characterize the relation $\lambda_{d\perp}=\lambda_{d\perp}(\lambda_{d\|},\mu_e)$, it can be derived from (\ref{traction1}) the family of stroke curves 
$\sigma_r=-\sigma_e(\lambda_{d\|}, \mu_e)$ which is shown} in figure \ref{fig:6} 
(solid lines, with $\lambda_c=\lambda_{d\|}$). 
The intercepts  of the curves with the vertical axis $\lambda_{c}=1$ yield the corresponding blocking forces:
\begin{equation}\label{bf1}
\sigma_r=-\sigma_e(1, \mu_e)= -G\,(\frac{1}{\lambda_{d\perp}^2}- 1)\,,
\,,
\end{equation}
being $\lambda_{d\perp}=\lambda_{d\perp}(1,\mu_e)$. Equation \eqref{bf1} shows \b{that} $\sigma_r=0$ for $\lambda_{d\perp}=1$; 
however, from being $\lambda_{d\perp}=\lambda_{d\perp}(1,\mu_e)$, it occurs that $\lambda_{d\perp}=1$ iff $\mu_e\to -\infty$. Hence, when constraints \b{maintain} the gel in the dry configuration under the special bath conditions characterised by $\mu_e\to -\infty$, blocking forces are null as that configuration is stress--free.

\b{Due to the anisotropic response, the characteristic stroke curves are} different when transverse--to--the--fibers--normal loads are considered.
We view $\sigma_e$ as the boundary traction exerted on the body by constraints hampering deformation along $\eb_2$ (see right inset in figure \ref{fig:6}), and  evaluate, for different values of $\mu_e$, 
the  traction corresponding to \b{a prescribed stretch  $\lambda_{d2}=\lambda_c$.} Precisely, using equation \eqref{num2} to characterise $\lambda_{d3}$ as $\lambda_{d3}(\lambda_{d1})$ and equation \eqref{case2} to characterise $\lambda_{d1}$ as $\lambda_{d1}(\lambda_{d2},\mu_e)$, the family of stroke curves $\sigma_r=-\sigma_e(\lambda_{d2})$ for different values of $\mu_e$ are shown in figure \ref{fig:6} (dashed lines, with $\lambda_c=\lambda_{d2}$).  The intercepts of the stroke curves with the vertical axis $\lambda_{c}=1$ define the  blocking forces for different $\mu_e$:
\begin{equation}\label{bf2}
\b{\sigma_r=-\sigma_{\eb_2} = \frac{G}{\lambda_{d1}}\left(\frac{1}{\lambda_{d3}}-\lambda_{d3}\right)\,,}
\end{equation}
\b{with $\lambda_{d3}=\lambda_{d3}(\lambda_{d1})$, and $\lambda_{d1}=\lambda_{d1}(1,\mu_e)$.} 
\b{In contrast to what happens in isotropic gels,} stroke--curves (dashed lines) are different from the ones discussed in the Section \ref{parallel} (solid lines), and the blocking forces exerted on the orthogonal faces of unit normals $\eb_1$ and $\eb_2$ are not the same, as figure \ref{fig:6} evidences (look at the intercepts of the stroke curves with the vertical axis $\lambda_{c}=1$).
\section{Anisotropic gels under tangential forces}
\b{Interestingly, the range of blocking forces generated by an anisotropic gel is quite large, 
as free--swelling can also induce shear, depending on the anisotropy directions.} 

\begin{figure}[h]
\centering
\includegraphics[scale=1.2]{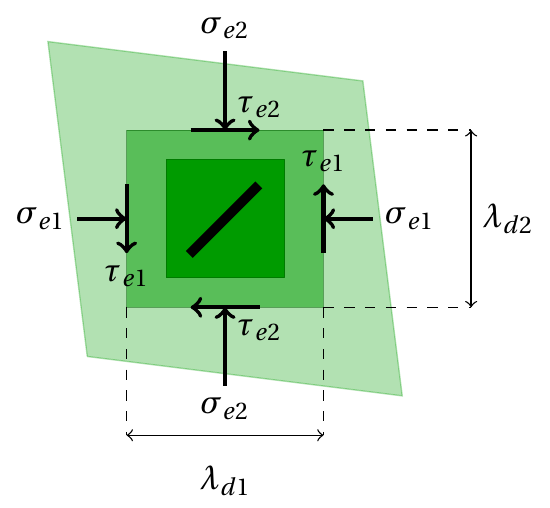}
\caption{Sketch of the plane section of unit normal $\eb_3$ of the unit cube at the dry state (light blue square), at the free--swelling state (light green diamond), and at the constrained state (green rectangle).}
\label{figF:10}
\end{figure}

Let us consider a fiber distribution within the hydrogel aligned along  $\eb=\sqrt{2}/2\eb_1 + \sqrt{2}/2\eb_2$, 
a situation that can \b{be easily generalized}. We imagine that appropriate constraints hamper the swelling in both the directions $\mb=\eb_1$ and $\mb=\eb_2$, so allowing the initial dry unit cube to swell into a parallelepiped of side $\lambda_{d1}$ and $\lambda_{d2}$, having in general $\lambda_{d1}\not=\lambda_{d2}$, 
and both of them smaller than $\lambda_{f\|}$, that is, 
\b{the linear swelling ratio along fiber's direction corresponding under a chemical potential $\mu_e$ to free swelling conditions} \b{(see figure \ref{figF:10}).} 
Hence, we assume that the representation \eqref{ciccio} of $\Fb_d$ still holds.
We look for homogeneous solutions of the stress--diffusion problem \b{posed on the dry configuration $\Bc_d$,
under the boundary conditions}
\begin{equation}\label{bal6}
\Sb_d\,\mb=\0
\,\,\text{for}\,\,\mb=\eb_3\quad\textrm{and}\quad
\mu=\mu_e\quad\text{on}\quad\partial\Bc_d\,,
\end{equation}
and prescribed \b{values of $\lambda_{d1}$ and $\lambda_{d2}$.} Hence,  mechanical and chemical balances prescribe
\begin{equation}\label{eqnBAL5}
s_{d3}=0\quad\textrm{and}\quad
\mu=\mu_e\qquad\textrm{in}\,\, \Bc_d\,,
\end{equation}
that is, using equations \eqref{csteqnani} and \eqref{bal6},  
\begin{eqnarray}
&& p = G\,\frac{\lambda_{d3}}{\lambda_{d1}\lambda_{d2}}\,,\label{case3}\\
&& \mu(J_d) + \Omega\,G\frac{\lambda_{d3}}{\lambda_{d1}\lambda_{d2}}=\mu_e\,,\quad J_d=\lambda_{d1}\lambda_{d2}\lambda_{d3}\,.\nonumber
\end{eqnarray}
\b{Equation \eqref{case3}$_2$ implicitly characterizes $\lambda_{d3}$ in terms of $\mu_e$ and the pair $(\lambda_{d1},\lambda_{d2})$, here considered as parameters}.

\begin{figure}[h]
\centering
\includegraphics[scale=.7]{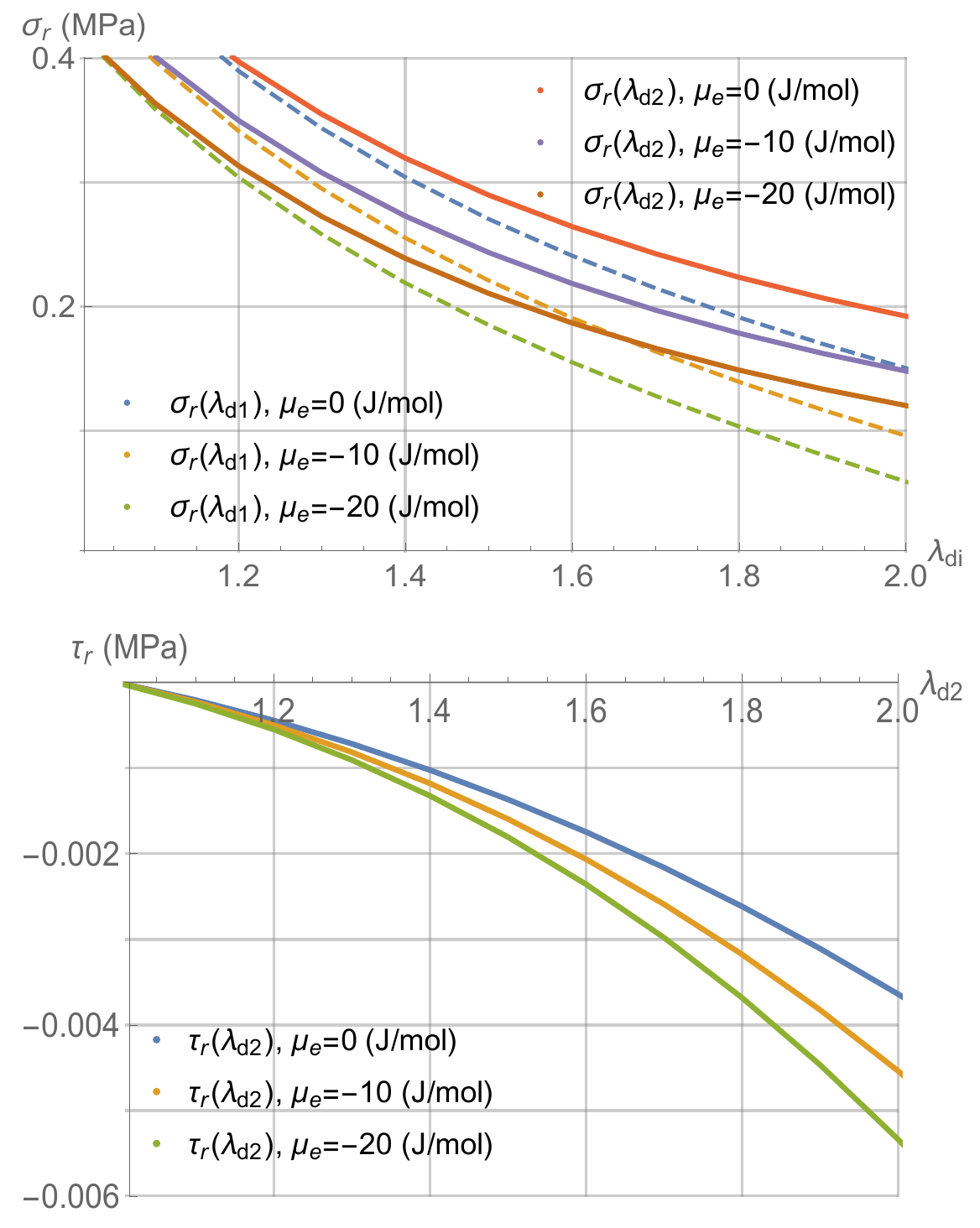}
\caption{Normal stroke curves  $\sigma_r(\lambda_{d1})$ at $\lambda_{d2})=1$ and $\sigma_r(\lambda_{d2})$ at $\lambda_{d1})=1$ and tangential stroke curves $\tau_r(\lambda_{d2})$ at $\lambda_{d1}=1$ at different values of $\mu_e$, being $G=0.1$MPa and $\gamma=0.1$.}
\label{figF:11}
\end{figure}

We consider the traction on the face of unit normal $\mb=\eb_1$; we have $\nb=\eb_1$ and fix $\tb=\eb_2$. With this, equations \eqref{tractions} prescribe 
\begin{eqnarray}
\sigma_{\eb_1} &=& G\Bigl(\frac{\lambda_{d1}}{\lambda_{d2}\,\lambda_{d3}}
\Bigl(1+\gamma\,\Bigl(\frac{\lambda_{d1}^2 +\lambda_{d2}^2}{2}-1\Bigr)\Bigr)
-\frac{\lambda_{d3}}{\lambda_{d1}\,\lambda_{d2}}\Bigr)\,,\nonumber\\
\tau_{\eb_1} &=& G\,\gamma\,\frac{1}{\lambda_{d3}}\Bigl(\frac{\lambda_{d1}^2 +\lambda_{d2}^2}{2}-1\Bigr)\,.\label{traction3}
\end{eqnarray}
\r{Due to the symmetry of the stress, tangential traction $\tau_{\eb_2}$ on the face  of unit normal $\mb=\eb_2$ (that is, for $\nb=\eb_1$) is equal to $\tau_{\eb_1}$, whereas in general $\sigma_{\eb_1}\not=\sigma_{\eb_2}$.} Using equations \eqref{case3}$_2$ and (\ref{traction3}), we can evaluate the normal stroke curves of the gel as $\sigma_r(\lambda_{d1})$ at $\lambda_{d2}=1$ and $\sigma_r(\lambda_{d2})$ at $\lambda_{d1}=1$,
 at different value of the solvent bath's potential $\mu_e$. Likewise, we can evaluate the tangential stroke curves $\tau_r(\lambda_{d2})$ at $\lambda_{d1}=1$. 
 It is worth noting that for $\lambda_{d1}=\lambda_{d2}=1$, it holds
\begin{equation}
\sigma_{\eb_1} = \sigma_{\eb_2}=G\Bigl(\frac{1}{\lambda_{d3}}-\lambda_{d3}\Bigr)\,,\label{traction4}
\end{equation}
and $\tau_{\eb_1} = \tau_{\eb_2}=0$. Figure \ref{figF:11} shows normal and tangential stroke curves corresponding to $G=0.1$ MPa, $\gamma=0.1$, and for different values of $\mu_e$; the range of $\lambda_{d2}$ goes from $1$, corresponding to dry conditions 
(being also $\lambda_{d1}=1$), to $\lambda_{d\parallel}$ (which is around $2.1$ with 
\b{the aforementioned values of $G$ and $\gamma$)}. Normal blocking forces are always positive, and tangential blocking forces are always negative, according to the cartoon shown in figure \ref{figF:10}.
\section{Conclusions}
We investigated  performances of anisotropic gels driven by mechanical and chemical stimuli, in terms of both deformation processes and stroke--curves. In some cases, we distinguished between the fast response of gels before-diffusion-starts, and the asymptotic response attained at the steady state, highlighting the difference in material relaxation due to diffusion. 

We also showed as anisotropic gel--based actuators can exert tangential, other than normal, blocking forces when fibers and constraints are not parallel and/or orthogonal each other. This kind of performances may be useful in actuator applications, which  have not been extensively studied when fibrous hydrogels are involved, even if Literature concerning technological applications is increasing.  
%
%%
%%%%%%%%%%%%%%%%%%%%%%%%%%%%%%%%%%%%%%%%%%% 
\begin{acknowledgments}
L.T.  acknowledges the National Group of Mathematical Physics (GNFM--INdAM) for support.
\end{acknowledgments}

%
%%
%\bibliography{rsc}
% Produces the bibliography via BibTeX.

\end{document}